\documentclass[a4paper,11pt]{article}
\usepackage{a4wide,amsmath,amssymb,bbm}
\usepackage[english]{babel}

\makeatletter\AtBeginDocument{\let\@elt\relax}\makeatother  

\usepackage{color}
\usepackage{slashed}
\usepackage[normalem]{ulem}
\usepackage{cite}
\usepackage[bottom]{footmisc}
\usepackage{graphicx}
\usepackage{adjustbox}
\usepackage{multirow}

\usepackage[pdfencoding=auto,psdextra]{hyperref}
\usepackage{enumerate}
\hypersetup{
    colorlinks,
    linkcolor={black},
    citecolor={black},
    urlcolor={black}
}

\makeatletter

\let\old@startsection=\@startsection
 \let\oldl@section=\l@section
 \renewcommand{\@startsection}[6]{\old@startsection{#1}{#2}{#3}{#4}{#5}{#6\mathversion{bold}}}
 \renewcommand{\l@section}[2]{\oldl@section{\mathversion{bold}#1}{#2}}
\makeatother

\makeatletter
\DeclareFontFamily{OMX}{MnSymbolE}{}
\DeclareSymbolFont{MnLargeSymbols}{OMX}{MnSymbolE}{m}{n}
\SetSymbolFont{MnLargeSymbols}{bold}{OMX}{MnSymbolE}{b}{n}
\DeclareFontShape{OMX}{MnSymbolE}{m}{n}{
    <-6>  MnSymbolE5
   <6-7>  MnSymbolE6
   <7-8>  MnSymbolE7
   <8-9>  MnSymbolE8
   <9-10> MnSymbolE9
  <10-12> MnSymbolE10
  <12->   MnSymbolE12
}{}
\DeclareFontShape{OMX}{MnSymbolE}{b}{n}{
    <-6>  MnSymbolE-Bold5
   <6-7>  MnSymbolE-Bold6
   <7-8>  MnSymbolE-Bold7
   <8-9>  MnSymbolE-Bold8
   <9-10> MnSymbolE-Bold9
  <10-12> MnSymbolE-Bold10
  <12->   MnSymbolE-Bold12
}{}

\let\llangle\@undefined
\let\rrangle\@undefined
\DeclareMathDelimiter{\llangle}{\mathopen}%
                     {MnLargeSymbols}{'164}{MnLargeSymbols}{'164}
\DeclareMathDelimiter{\rrangle}{\mathclose}%
                     {MnLargeSymbols}{'171}{MnLargeSymbols}{'171}
\makeatother

\usepackage{mathrsfs}

\usepackage{comment}
\usepackage{braket}

\newcommand{\STr}{\mathrm{STr}}
\newcommand{\be}{\begin{equation}}
\newcommand{\ee}{\end{equation}}
\newcommand{\alg}{\mathfrak}

\newcommand{\AD}{\operatorname{Ad}}

\newcommand{\g}{\mathfrak{g}}
\newcommand{\h}{\mathsf{h}}
\newcommand{\e}{\mathsf{e}}
\newcommand{\Q}{\mathsf{Q}}
\newcommand{\as}{\mathsf{a}}
\newcommand{\bs}{\mathsf{b}}
\newcommand{\is}{\mathsf{i}}
\newcommand{\js}{\mathsf{j}}
\newcommand{\Is}{\mathsf{I}}
\newcommand{\Js}{\mathsf{J}}
\newcommand{\T}{\mathsf{T}}
\newcommand{\ts}{\mathsf{t}}

\newcommand{\RB}[1]{{\color[rgb]{0,0,1} #1}}

\makeatletter
\renewcommand{\fnum@figure}[1]{\textbf{\footnotesize Figure~\thefigure :}}

\@addtoreset{equation}{section} \makeatother

\begin{document}

\null

\vspace{30pt}

\begin{center}
{\huge{\bf  All Jordanian deformations of\\\vspace{.4cm}
the $AdS_5 \times S^5$ superstring}}

\vspace{80pt}

Riccardo Borsato, \ \ Sibylle Driezen

\vspace{15pt}

{
\small {{\it 
Instituto Galego de F\'isica de Altas Enerx\'ias (IGFAE) and
Departamento de F\'\i sica de Part\'\i culas,\\[7pt]
Universidade de  Santiago de Compostela\\
\vspace{12pt}
\texttt{riccardo.borsato@usc.es, sib.driezen@gmail.com}}}}\\

\vspace{100pt}

{\bf Abstract}
\end{center}
\noindent
We explicitly construct and classify all  Jordanian solutions of the classical Yang-Baxter equation on $\mathfrak{psu}(2,2|4)$, corresponding to Jordanian Yang-Baxter deformations of the $AdS_5\times S^5$ superstring. Such deformations preserve the classical integrability of the underlying sigma-model and thus are a subclass of all possible integrable deformations. The deformations that we consider are divided into two families, unimodular and non-unimodular ones. The former ensure that the deformed backgrounds are still  solutions of the type IIB supergravity equations. For the simplest unimodular solutions, we find that the corresponding backgrounds preserve a number $N<32$ of supercharges  that can be $N=12,8,6,4,0$.

\pagebreak

\setcounter{page}{1}
\newcounter{nameOfYourChoice}

\tableofcontents

\section{Introduction}

Recent years have seen an upsurge in the understanding of integrable deformations of two-dimensional sigma-models, including their possible classifications and applications. Well-known cases are TsT-transformations \cite{Lunin:2005jy,Frolov:2005dj,Frolov:2005ty}, Yang-Baxter deformations \cite{Klimcik:2002zj,Klimcik:2008eq,Delduc:2013fga,Delduc:2013qra,Kawaguchi:2014qwa,vanTongeren:2015soa}, and $\lambda$-deformations \cite{Sfetsos:2013wia,Hollowood:2014rla,Hollowood:2014qma}. Major motivations for their study follow from developing a fundamental understanding of integrable two-dimensional field theories, their relation with generalised worldsheet dualities, and their prospect in generalising the AdS/CFT correspondence to non-maximally supersymmetric cases whilst preserving the computational power of integrability (see e.g.~the reviews \cite{Thompson:2019ipl,Hoare:2021dix}). In particular, when the sigma-model describes the dynamics of a string then an integrable deformation on its worldsheet will deform the  target-space background destroying many of its (super)isometries. Whether or not the deformed background will still solve the supergravity equations of motion is now well-understood for large classes of deformations (see e.g.~\cite{Borsato:2016ose,Hoare:2018ngg,Sfetsos:2014cea,Appadu:2015nfa,Borsato:2021gma,Borsato:2021vfy}). This thus opens exciting possibilities for applications in particular  to the canonical example of the $AdS_5\times S^5$ string, with hopes to find deformations of ${\cal N}=4$ super-Yang-Mills that remain integrable in the planar limit~\cite{Beisert:2010jr}.

In this paper, we will focus on a particular type of deformations, called Jordanian, which belong to the homogeneous deformations of the Yang-Baxter class \cite{Kawaguchi:2014qwa,vanTongeren:2015soa}. Homogeneous deformations are generated by a linear operator $R$ that acts on the Lie (super)algebra of (super)isometries $\mathfrak{g}$ of an integrable sigma-model, and they preserve integrability if $R$ is antisymmetric as in eq.~\eqref{eq:antisymmR} and satisfies the classical Yang-Baxter equation (CYBE) given in eq.~\eqref{eq:CYBE}.\footnote{For the inhomogeneous version, the antisymmetric $R$ must instead solve the modified Yang-Baxter equation  \cite{Klimcik:2002zj,Klimcik:2008eq,Delduc:2013fga,Delduc:2013qra}.}  The CYBE admits a rich number of antisymmetric solutions, all of which will lead to different deformed target-space backgrounds of the string sigma-model.
These backgrounds will be supergravity solutions when the $R$-matrix is unimodular \cite{Borsato:2016ose,Borsato:2017qsx,Borsato:2018idb}, which is a  simple linear constraint on $R$, see eq.~\eqref{eq:uni-cond}. When instead $R$ is non-unimodular, they will solve the modified (or generalised) supergravity equations identified in \cite{Arutyunov:2015mqj,Wulff:2016tju}. 
Within the homogeneous Yang-Baxter models, $R$-matrices of Jordanian type  are built by identifying a bosonic subalgebra of $\mathfrak{g}$ constructed from a Cartan element $\h$ and a root $\e$ satisfying $[\h,\e]=\e$. When $\h$ and $\e$ are the only elements in its construction, the Jordanian $R$-matrix is of rank-2 and will in fact be non-unimodular. However, it was found in \cite{vanTongeren:2019dlq} that at least some of these rank-2 cases can be extended to unimodular $R$-matrices by employing, next to $\h$ and $\e$, also fermionic generators (supercharges) of the superalgebra $\mathfrak{g}$ in its construction.

Our focus on the homogeneous deformations of Jordanian type originates from several motivations. In general, homogeneous deformations  generalise the well-known TsT (T-duality-shift-T-duality) transformations \cite{Lunin:2005jy,Frolov:2005dj,Frolov:2005ty}  to the case where the subalgebra participating in the construction of $R$ is non-abelian. See~\cite{Osten:2016dvf} for the reformulation of TsT transformations as Yang-Baxter deformations. For TsT, this subalgebra is abelian and corresponds to the commuting isometries along which the T-dualities are performed. An important property of TsT is that the corresponding deformed models with periodic worldsheet boundary conditions can be reformulated as undeformed models with twisted worldsheet boundary conditions \cite{Frolov:2005dj,Frolov:2005ty,Alday:2005ww} that are local.  In the case of ``diagonal’’ TsT-models, where the object causing the twisting is diagonal, this is crucial in the understanding of their spectral problem  by means of integrability methods \cite{Beisert:2005if,deLeeuw:2012hp,Kazakov:2018ugh}.\footnote{From the point of view of the holographic duality, understanding these deformations as deformations of the super Yang-Mills action is not always obvious, in particular when on the string side of the correspondence the AdS space participates in the deformation. We refer to~\cite{Meier:2023kzt} for recent developments in this direction.}
Recently, the reformulation in terms of a (local) twisted model has been achieved for  generic homogeneous Yang-Baxter deformations \cite{Borsato:2021fuy}.\footnote{For earlier work giving a reformulation in terms of undeformed models with non-local twisted boundary conditions see \cite{Matsumoto:2015jja,Vicedo:2015pna,vanTongeren:2018vpb}.} In contrast to other options, it was found that the Jordanian $R$-matrices lead to a twist that is always diagonalisable, and therefore one could hope to apply similar integrability techniques that worked for TsT to tackle their spectral problem. This has been done successfully at the semi-classical level in \cite{Borsato:2022drc} for a specific Jordanian deformation of $AdS_5\times S^5$ (which corresponds to our $R_1$ below with $a=0,b=-1/2$ and to $\bar R_1$ with $a=0,b=-1/2,q_{2,22}^+=0$ for its unimodular version).

In this paper, we will continue the study of string sigma-models deforming the canonical $AdS_5\times S^5$ background\footnote{Although our main interest is in $AdS_5\times S^5$, some of our results are useful to classify also deformations of backgrounds $AdS_n\times M$ with $n<5$. In fact some of our $R$-matrices are constructed with generators of $\alg{so}(2,4)$ that are also elements  of the isometry groups of $AdS_n, n<5$, and therefore they can be used to generate corresponding deformations.} and therefore take $\mathfrak{g}=\mathfrak{psu}(2,2|4)$. To diversify the applications of Jordanian models,  we will classify all antisymmetric solutions to the CYBE that are of Jordanian type.  This includes a classification of all canonical rank-2 Jordanian $R$-matrices as well as their bosonic extensions. We will find that they are at most of rank-6. For all of the bosonic $R$-matrices, we will explicitly construct only those fermionic extensions that ensure unimodularity and thus a corresponding supergravity background that is well-behaved. We will pay particular attention to simplifying our results as much as possible, including the  identification of  equivalent solutions, by means of inner automorphisms of $\mathfrak{g}=\mathfrak{psu}(2,2|4)$.

The paper is organised as follows. We present a summary of our main results in section \ref{sec:summary}. In section \ref{sec:jord}, we define Jordanian $R$-matrices including their bosonic and fermionic extensions, as well as the unimodularity condition. This in particular identifies the necessary requirements that the subalgebra used to construct the $R$-matrix has to satisfy. In section \ref{sec:rank2}, we present our classification of the bosonic rank-2 $R$-matrices. These results are summarised in Table \ref{tab:rank-2-un} and \ref{tab:rank-2-no-un}. In section \ref{sec:bos-ext}, we will construct their bosonic higher-rank extensions---which only exist for certain rank-2 cases---summarised in the rest of section~\ref{sec:summary}. In section \ref{sec:uni-ext}, we then construct all the fermionic extensions that ensure unimodularity. Again in many of the rank-2 cases, we will see that they actually do not admit a unimodular extension, while all higher-rank cases do. In section \ref{sec:siso}, we revive our string theory motivation, and identify for all possible unimodular extensions of the rank-2 $R$-matrices the number of (super)isometries that are preserved in the corresponding deformed supergravity background.
The number of preserved superisometries  is summarised in Table \ref{tab:rank-2-siso}, while the inclusion of bosonic isometries is presented in Table \ref{tab:rank-2-siso-full}. We end with some conclusions in section \ref{sec:conclusions}. Appendix \ref{app:conv} collects facts and conventions for homogeneous $R$-matrices and Yang-Baxter deformations, while appendix \ref{app:psu224} collects our conventions for the $\mathfrak{psu}(2,2|4)$ algebra and presents the explicit matrix realisation that we used.

\section{Summary of the results} \label{sec:summary}
For the reader's convenience, in this section we present a summary of our main results.  All our results are presented modulo inner automorphisms of the algebra, or modulo transformations that leave the $R$-matrix invariant. Inner automorphisms  of the algebra map equivalent solutions of the $R$-matrices to each other: from the point of view of the sigma model the two deformations would be related by field redefinitions, and from the point of view of the 10-dimensional background   by coordinate transformations. 

The bosonic Jordanian $R$-matrices may be grouped by their rank. We find that we can have bosonic Jordanian solutions of rank 2, 4 and 6. In the following we present the results for the bosonic $R$-matrices and their unimodular extensions.

\subsection*{Rank-2} 
The simplest possible case is that of rank 2, with $R$-matrices of the form $r=\h\wedge \e$. Here we are using a notation to write the $R$-matrix that is reviewed in appendix~\ref{app:conv}. In Tables~\ref{tab:rank-2-un} and~\ref{tab:rank-2-no-un} we collect all possible rank-2 $R$-matrices of $\alg{so}(2,4)$. In Table~\ref{tab:rank-2-un} we list the ones that admit a unimodular extension, while in Table~\ref{tab:rank-2-no-un} the ones that do not admit it.
It turns out that only $R$-matrices with $\e=p_0+p_3$ or $\e=p_0$ admit a unimodular extension.

\begin{table}[h!]
\centering
\begin{adjustbox}{center}
    \begin{tabular}{c|c|c|c}
    $R$ & $\h$ & $\e$ & residual $\mathrm{Inn}(\mathfrak{so}(2,4))$ \\
    \hline       
            &  &  & $J_{12}$, $D+J_{03}$  (and $p_1,p_2$ if $a=0$ and $b=-1$;\\ 1 & $(1+b)D+b J_{03}+a J_{12}$ & $p_0+p_3$ &  and $J_{01}-J_{13},J_{02}-J_{23}$ if $a=b=0$;  \\
            & & & and $p_0-p_3, k_0+k_3$ if $b=-1/2$)\\ 
            \hline    
    2& $\tfrac12(D-J_{03})+a J_{12}+\alpha(p_0-p_3)$  & $p_0+p_3$ & $J_{12}, p_0 - p_3$ \\
    \hline    
    3& $-J_{03}+\alpha p_1$  & $p_0+p_3$ & $p_1,p_2$ \\
    \hline    
    4& $D+\alpha(J_{01}-J_{13})$ & $p_0+p_3$ & $J_{01}-J_{13},J_{02}-J_{23}$ \\
    \hline    
    5& $\tfrac12(D-J_{03})+a J_{12}+b (k_0+k_3+2p_3)$ & $p_0+p_3$ & $J_{12}$, $k_0+k_3-p_0+p_3$ \\
    \hline    
    6& $D+a J_{12}$ & $p_0$ & $J_{12}$  (and $J_{13},J_{23}$ if $a=0$) \\
    \end{tabular}
\end{adjustbox}
    \caption{Rank-2 bosonic Jordanian deformations of the form $r=\h\wedge \e$ that admit a unimodular extension. The convention is that the parameter $\alpha$ squares to 1 $(\alpha^2=1)$ and $a,b \in \mathbb{R}$ are free. In the last column we write the residual inner automorphisms in $\mathfrak{so}(2,4)$ that (together with $\mathfrak{so}(6)$) leave $\h$ and $\e$ invariant.  In those lists we omit $\e$ itself, which always corresponds to an isometry.
    }
    \label{tab:rank-2-un}
\end{table}

\begin{table}[h!]
    \centering
    \begin{tabular}{c|c}
    $\h$ & $\e$ \\
    \hline
         $D+a J_{03}$ & $p_1$ \\
         \hline    
   $D+\alpha(J_{02}-J_{23})$ & $p_1$ \\
   \hline    
   $D+\alpha(p_0+\beta p_3)+\beta J_{03}$  & $p_1$ \\
   \hline    
     $-J_{03}+a D$ & $J_{02}-J_{23}$ \\
     \hline    
     $-J_{03}+\alpha p_1$ & $J_{02}-J_{23}$ \\
     \hline    
     $-J_{03}-D+\alpha(p_0+ p_3)$ &  $J_{02}-J_{23}$\\
     \hline    
        $D-J_{03}$ & $ap_1+bp_2+J_{01}-J_{13}$\\
        \hline    
       $D+a J_{12}$ & $p_3$\\
       \hline    
     $2D-J_{03}$ & $p_0-p_3+J_{01}-J_{13}$  \\
     \hline    
    $D-J_{03}- 2a\alpha J_{12}+a (k_0+k_3+2p_3)$ & $\alpha p_2+J_{01}-J_{13}$ 
    \end{tabular}
    \caption{Rank-2 bosonic Jordanian deformations of the form $r=\h\wedge \e$ that do \emph{not} admit a unimodular extension. The convention is that the parameters $\alpha,\beta$ square to 1, $(\alpha^2=\beta^2=1)$, and $a,b \in \mathbb{R}$ are free.}
    \label{tab:rank-2-no-un}
\end{table}
In the last column of Table~\ref{tab:rank-2-un} we list the inner automorphisms of $\alg{so}(2,4)$ that are residual isometries under the deformation and that leave $\h$ and $\e$ invariant. It turns out that this information is useful also to simplify as much as possible the unimodular extensions of these rank-2 solutions. In particular, a unimodular $R$-matrix will be of the form
\begin{equation}
    r=\h\wedge\e-\frac{i}{2}(\Q_1\wedge\Q_1+\Q_2\wedge\Q_2),
\end{equation}
where $\Q_{\is},\ \is=1,2$ are odd elements of $\alg{psu}(2,2|4)$ of the form
\begin{equation} 
    \Q_\is=q^{+}_{\is,\alpha a} \mathcal Q_+^{\alpha a} +i\, q^{-}_{\is,\alpha a} \mathcal Q_-^{\alpha a},
\end{equation}
where $ \mathcal Q_\pm^{\alpha a}\equiv \overline Q^{\alpha a}\pm \epsilon^{\alpha\beta}Q_{\beta a}$ and $q_{\is,\alpha a}^\pm$ are real numbers. We refer to appendix~\ref{app:psu224} for our conventions for $\alg{psu}(2,2|4)$. 

It turns out that the unimodular extensions of rank-2 solutions come in four types: 
\begin{enumerate}
    \item First we have $\bar R_1, \bar R_2, \bar R_3, \bar R_4,\bar R_5$ as unimodular extensions of the corresponding bosonic $R$-matrices $R_1,R_2,R_3,R_4,R_5$. This type of extension works only if there is no $J_{12}$ in $\h$, so that one has to set $a=0$ in $R_1,R_2,R_5$. We find that we can set to zero all coefficients $q_{\is,\alpha a}^\pm$ except three of them that satisfy
\begin{equation}\label{eq:constrqRbar1}
    q^+_{1,21} = \frac{1}{\sqrt{2}}, \qquad  (q^+_{2,22})^2+(q^-_{2,21})^2=\frac{1}{2} \ .
\end{equation}
Notice that this solution admits one continuous parameter. It is a physical parameter, in fact when $q^+_{2,22}=0$ there is an enhancement of the number of superisometries (see below). 

\item We then have the second type of extension, that can be constructed when there is a non-trivial contribution of $J_{12}$ in $\h$. That means that we assume $a\neq 0$ in $R_1,R_2,R_5$ and construct the unimodular $R$-matrices $\bar R_{1'},\bar R_{2'},\bar R_{5'}$. In this case all coefficients $q_{\is,\alpha a}^\pm$ can be set to zero except two of them
  \begin{equation} 
        q^-_{2,21}=q^+_{1,21}=\frac{1}{\sqrt{2}}  \ .
    \end{equation}
    Taking the $a\to 0$ limit one recovers the previous type of extension with the extra condition $q^+_{2,22}=0$.

\item The third type of extension will be denoted by $\bar R_6$, which is the unimodular extension of $R_6$ when there is no $J_{12}$ contribution in $\h$ (i.e.~$a=0$). We find that the coefficients must satisfy
\begin{equation}
    \begin{aligned}
    & q^-_{1,11}=q^-_{1,22}=\frac{1}{2},  \qquad q^-_{2,21}=- q^-_{2,12}, \\
        &(q^+_{2,22})^2 +(q^+_{2,23})^2 + (q^+_{2,24})^2 + (q^-_{2,21})^2 + (q^-_{2,24})^2 = \frac{1}{4} , \\
        &(q^+_{2,11})^2 + (q^+_{2,13})^2 + (q^-_{2,12})^2 + (q^-_{2,13})^2 = \frac{1}{4} , \\
        & (q^+_{2,11}+q^+_{2,22}) q^-_{2,21} - q^-_{2,13} q^+_{2,23} =0 , \\
        &  q^+_{2,13} q^+_{2,23} =0 . 
    \end{aligned}
\end{equation}
while all other coefficients can be set to zero.

\item Finally, we have the extension $\bar R_{6'}$ of $R_6$ in the case $a\neq 0$. Now all coefficients are zero except
\begin{equation}
    q^-_{1,11}=q^+_{2,11}=   q^-_{1,22}=- q^+_{2,22}=\frac{1}{2}  \ .
\end{equation}

\end{enumerate}

In Table~\ref{tab:rank-2-siso} we summarise the number of superisometries that are preserved by the deformation for each of the unimodular extensions of the rank-2 $R$-matrices. A table that includes also the residual bosonic isometries is given in the main text, see Table \ref{tab:rank-2-siso-full}.

\begin{table}[h!]
\centering
\begin{adjustbox}{center}
    \begin{tabular}{l|c|c}
    $\bar{R}$ & \multicolumn{1}{|c|}{conditions}  & supercharges  \\
    \hline

    \hline
    \multirow{3}{*}{1} & ${a=0}$ &   0\\
    \cline{2-3}
     &  ${a=0}, b=-1/2$   &    8\\
    \cline{2-3}
     &  ${a=0}, b=-1/2$, $q^+_{2,22}=0$   &     12\\
    \hline

    \hline
    \multirow{1}{*}{1'} & $a\neq 0$ &   0 \\
    \hline
    
    \hline
    \multirow{2}{*}{2}
     & ${a=0}$ &   4 \\
    \cline{2-3}
     &  ${a=0}, q^+_{2,22}=0$ &   6 \\
    \hline

    \hline
    \multirow{1}{*}{2'} & $a\neq 0$ &   0 \\
    \hline

    \hline
    \multirow{1}{*}{3} & $-$ &  0\\
    \hline

    \hline
    \multirow{1}{*}{4} & $-$ & 0 \\
    \hline

    \hline
    \multirow{1}{*}{5} & ${a=0}$ & 0 \\
    \hline

    \hline
    \multirow{1}{*}{5'} & $a\neq 0$ & 0 \\
    \hline

    \hline
    6 & $a=0$&  0\\
  \hline
     6' & $a\neq 0$&  0
    \end{tabular}
\end{adjustbox}
    \caption{The number of independent supercharges $T_{\bar{A}}\in \mathfrak{psu}(2,2|4)$ that satisfy $\mathrm{ad}_{T_{\bar{A}}} R = R \mathrm{ad}_{T_{\bar{A}}}$ for  the unimodular extensions of the rank-2 Jordanian $R$-matrices. They have the form $r=\h\wedge \e -\frac{i}{2} (\Q_1\wedge \Q_1 + \Q_2 \wedge \Q_2)$. Such elements represent superisometries of the deformed supergravity background. If a parameter is not specified, it is assumed to be generic (modulo constraints such as \eqref{eq:constrqRbar1}). }
    \label{tab:rank-2-siso}
\end{table}

\subsection*{Rank-4}
Higher-rank bosonic Jordanian $R$-matrices can be constructed only in the case of $\e=p_0+p_3$. 
Rank-4 solutions are of the form $r=\h\wedge\e+\e_+\wedge\e_-$, they are all listed in Table~\ref{tab:rank-4} and they all admit a unimodular extension.

\begin{table}[h!]
    \centering
    \begin{tabular}{c|c|c|c}
    $R$ & $\h$& $\e_+$& $\e_-$ \\
    \hline
   $7$ & $(1+b)D+b\, J_{03}$  & $a\, p_1+  p_2$ & $ j_{023}$ \\
   \hline
       $8$ & $\tfrac23 D-\tfrac13 J_{03}$ & $a\,  p_1+  p_2$ & $b\, (p_0-p_3)+ j_{023}$ \\
         \hline            $9$ & $\tfrac12( D- J_{03})$ & $\, p_1+b_+p_2+d_+ j_{023}$ & $\begin{array}{c}b_-p_2 +d_-j_{023}\\
            +(1-b_+d_-+b_-d_+) j_{013}\end{array}$ \\
            \hline
  $10$ & $-J_{03}+\alpha\,  p_1$  & $- \, j_{023}$ & $a\,  p_1+ p_2$ \\
      \hline
    $11$ & $D+\alpha j_{013}$  & {$ p_2$} & $a j_{013}+  j_{023}$ \\
                 \hline
          $12$    & $\begin{array}{c}\tfrac12(D-J_{03})+a J_{12}\\+\frac{\alpha a}{2} (k_0+k_3+2p_3)\end{array}$ 
             & $p_1+\alpha \, j_{023}$
             & $
            - \tfrac{1}{2}(\alpha p_2- j_{013})$
             
    \end{tabular}
    \caption{Rank-4 bosonic Jordanian deformations of the form $r=\h\wedge\e+\e_+\wedge\e_-$. In all these cases we have $\e=p_0+p_3$. To save space we are using the shorthand notation $j_{\mu\nu\rho}=J_{\mu\nu}-J_{\nu\rho}$. All these $R$-matrices admit unimodular extensions, with no further restriction of the above parameters. When we have the parameter $\alpha$, it is assumed that $\alpha^2=1$, while all other parameters with latin letters are free real numbers. }
    \label{tab:rank-4}
\end{table}

The unimodular extensions of these $R$-matrices are
constructed with 4 (rather than just 2) odd elements of $\alg{psu}(2,2|4)$
\begin{equation}
    r=\h\wedge\e+\e_+\wedge\e_--\frac{i}{2}\sum_{\Is=1,2}(\Q^\Is_1\wedge\Q^\Is_1+\Q^\Is_2\wedge\Q^\Is_2),
\end{equation}
where we added a label $\Is=1,2$ and we still take
\begin{equation} 
    \Q^\Is_\is=q^{+,\Is}_{\is,\alpha a} \mathcal Q_+^{\alpha a} +i\, q^{-,\Is}_{\is,\alpha a} \mathcal Q_-^{\alpha a}.
\end{equation}
The unimodular extensions of $R_7,R_8,R_9,R_{10},R_{11}$ are of the same type and can be taken as follows.
The coefficients for $\Q^1_1$ and $\Q^1_2$ can be simplified as in the extension $\bar R_1$, meaning that we can set
\begin{equation}\label{eq:simpl-Q1}
    q^{+,1}_{1,21} = \frac{1}{\sqrt{2}}, \qquad  (q^{+,1}_{2,22})^2+(q^{-,1}_{2,21})^2=\frac{1}{2} \ .
\end{equation}
For $\Q_1^2$ and $\Q_2^2$ we can have in principle 7 non-trivial parameters turned on, satisfying the conditions~\eqref{eq:uni-ext-rank4and6-1} and with the extra choice of $q^{\pm,2}_{1,24}=q^{-,2}_{1,23}=q^{+,2}_{\is,21}=0$. 

The result for the unimodular extension of $R_{12}$ is simpler to present, because all coefficients can be set to zero except
\begin{equation}
    q^{-,1}_{1,21}=-q^{+,1}_{2,21}=  q^{-,2}_{1,22} =-q^{+,2}_{2,22}= \frac{1}{\sqrt{2}} \ .
\end{equation}
Notice that there is no continuous parameter left in the odd elements used to construct this $R$-matrix.

\subsection*{Rank-6}
There are two possible types of rank-6 bosonic solutions. They are of the form
$r=\h\wedge\e+\e_{+1}\wedge\e_{-1}+\e_{+2}\wedge\e_{-2}$
and in both cases
\begin{equation}
        \h=(1+b)D+b J_{03},\qquad \e=p_0+p_3.
\end{equation}
The first option is to take
\begin{equation}
    R_{13}:\quad
     \e_{+1}=a p_1+ p_2,\qquad 
     \e_{-1}=J_{02}-J_{23},\qquad 
     \e_{+2}=p_1,\qquad 
     \e_{-2}=J_{01}-J_{13}- a(J_{02}-J_{23}).
\end{equation}
The second option is to set
\begin{equation}
    R_{14}:\quad 
    b=-\frac12,\qquad 
    \e_{\pm \as}=a_{\pm\as }p_1+b_{\pm\as }p_2+c_{\pm\as }\, j_{013}+d_{\pm\as }\, j_{023},
\end{equation}
while imposing the  conditions
\begin{equation}
    \begin{aligned}
         a_{\pm 1}c_{\pm 2}+b_{\pm 1}d_{\pm 2}
       -b_{\pm 2}d_{\pm 1}-a_{\pm 2}c_{\pm 1}&=0,  \\
          a_{\pm 1}c_{\mp 2}+b_{\pm 1}d_{\mp 2}
          -b_{\mp 2}d_{\pm 1}-a_{\mp 2}c_{\pm 1}&=0,\\
          a_{+\as }c_{-\as }+b_{+\as }d_{-\as }
          -a_{-\as }c_{+\as }-b_{-\as }d_{+\as }&=1,\quad \as =1,2.
    \end{aligned}
\end{equation}
The unimodular extensions of rank-6 solutions are constructed with 6 odd elements of the superalgebra
\begin{equation}
r=\h\wedge\e+\e_{+1}\wedge\e_{-1}+\e_{+2}\wedge\e_{-2}-\frac{i}{2}\sum_{\Is=1}^3(\Q^\Is_1\wedge\Q^\Is_1+\Q^\Is_2\wedge\Q^\Is_2).
\end{equation}
The coefficients must again solve the constraints given in~\eqref{eq:uni-ext-rank4and6-1}. For $\Q^1_\is$ we may again simplify the solution to~\eqref{eq:simpl-Q1}, and for $\Q^2_\is$ we can set $q^{\pm,2}_{1,24}=q^{-,2}_{1,23}=q^{+,2}_{\is,21}=0$. We were not able to simplify the other coefficients further.

\subsection*{Shifts by $\mathfrak{so}(6)$}
So far we have presented the bosonic Jordanian solutions of the $\mathfrak{so}(2,4)$ algebra and their unimodular extensions in $\mathfrak{psu}(2,2|4)$. However, the full bosonic subalgebra of $\mathfrak{psu}(2,2|4)$ is $\mathfrak{so}(2,4)\oplus \mathfrak{so}(6)$, and this permits the construction of slightly more general Jordanian solutions. In particular, starting from the $\mathfrak{so}(2,4)$ solutions of rank $2,4,6$, we can shift $\h\to\h'=\h+\ts$ where $ \ts=\nu_{1}\, \mathcal R_{12}+\nu_{2}\, \mathcal R_{35}+\nu_{3}\, \mathcal R_{46}$ and $ \mathcal R_{12},\mathcal R_{35},\mathcal R_{46}\in \mathfrak{so}(6)$ are the three Cartan generators. This covers all the possible (extended) bosonic Jordanian solutions in $\mathfrak{so}(2,4)\oplus \mathfrak{so}(6)$. 

Given that $\e$ always generates an isometry in the presence of a Jordanian deformation, these solutions with $\mathfrak{so}(6)$ shifts can be interpreted as the composition of a Jordanian deformation followed by a TsT deformation along the $\ts,\e$ directions.\footnote{See e.g.~section 3.1 of~\cite{Borsato:2018spz} for the composition of compatible Yang-Baxter deformations.}

To construct unimodular extensions of the shifted Jordanian $R$-matrices we can start from the unimodular extensions of the $\mathfrak{so}(2,4)$ $R$-matrices and impose extra conditions. The options are the following. If we want the parameters $\nu_1,\nu_2,\nu_3\in \mathbb{R}$ to be generic, then we must further impose
\begin{equation}    q^{+,\Is}_{\is,\alpha a}=\epsilon_{\is\js}q^{-,\Is}_{\js,\alpha a},\qquad \forall\ \Is,\is,\alpha,a\quad\text{and}\quad
q^{+,\Is}_{\is,\alpha 2}=q^{+,\Is}_{\is,\alpha 3}=q^{+,\Is}_{\is,\alpha 4}=0.
\end{equation}
Notice that in some cases the unimodular extensions of the $\mathfrak{so}(2,4)$ $R$-matrices already satisfy this condition.
One can relax the conditions $q^{+,\Is}_{\is,\alpha 2}=q^{+,\Is}_{\is,\alpha 3}=0$ by turning on also coefficients with $a=2$ or $a=3$ but that  would   restrict the possible values of the parameters  $\nu_1,\nu_2,\nu_3$. 

For special values of the parameters $\nu_1,\nu_2,\nu_3$ the  condition $q^{+,\Is}_{\is,\alpha a}=\epsilon_{\is\js}q^{-,\Is}_{\js,\alpha a}$ can be dropped, but other conditions must be imposed. Up to automorphisms, there are two options: if $\nu_3=0,\nu_1=-\nu_2$ then $q^{+,\Is}_{\is,\alpha 3}=q^{+,\Is}_{\is,\alpha 4}=0$;  if $\nu_3=-\nu_1-\nu_2\neq 0$ then $q^{+,\Is}_{\is,\alpha 2}=q^{+,\Is}_{\is,\alpha 3}=q^{+,\Is}_{\is,\alpha 4}=0$.

\section{Extended Jordanian $R$-matrices}\label{sec:jord}

Given a Lie superalgebra $\g$, we identify a Cartan element $\h$ and the generator $\e$ of a positive root. Both of them will be of even grading (i.e.~$\text{deg}(\h)=\text{deg}(\e)=0$), in other words they belong to a standard Lie algebra. The assumption is that $\h$ and $\e$ span a subalgebra of $\g$ with
\begin{equation}
[\h,\e]=\e.    
\end{equation}
When this is the case, they identify a so-called ``Jordanian solution'' of the classical Yang-Baxter equation that is given by
\begin{equation}
    r=\h\wedge\e.
\end{equation}
We refer to appendix~\ref{app:conv} for our conventions on $R$-matrices.

Given a Jordanian $R$-matrix as above, it is possible to construct ``extended Jordanian'' solutions following Tolstoy~\cite{2004Tolstoy}. The extra ingredients are $N$ pairs of generators in $\alg g$ that we will denote as $\{\e_{i},\e_{-i}\}$ with $i=1,\ldots,N$, where $\e_{i}$ (resp.~$\e_{-i}$) corresponds to a positive (resp.~negative) root. These extra generators may be of even or odd grading, but they must satisfy $\deg(\e_{i})=\deg(\e_{-i})$. Moreover, the following  graded commutation relations must hold\footnote{Compared to~\cite{2004Tolstoy} here we use the parameter $\xi$ which is related to the parameter $t$ as $t=\tfrac12-\xi$.}
\begin{equation}\label{eq:ext-rel-Tol}
\begin{aligned}
\relax [ \e_{\pm i},\e ]=0,\qquad
[\h,\e_{\pm i}]=(\tfrac12\pm \xi_i)\e_{\pm i},\qquad
[[\e_k,\e_l]]=\delta_{k,-l}\, \e,
\end{aligned}
\end{equation}
with $k>l\in \{\pm 1,\pm 2,\ldots, \pm N\}$ and $ \xi_i\in \mathbb C$. 
Here $[[,]]$ denotes the graded commutator, see appendix~\ref{app:conv}.
The $N$-extended Jordanian $R$-matrix is then constructed as
\begin{equation}\label{eq:ext-r-Tol}
r=\h\wedge \e -\sum_{i=1}^N\e_{-i}\wedge \e_{i},
\end{equation}
where 
we used the graded wedge product\footnote{The definition implies that if $\mathsf a,\mathsf b$ are even then $\mathsf a\wedge \mathsf b=-\mathsf b\wedge \mathsf a$ while if they are odd $\mathsf a\wedge \mathsf b=+\mathsf b\wedge \mathsf a$.}
\begin{equation}
\mathsf a\wedge \mathsf b=\mathsf a\otimes \mathsf b-(-1)^{\text{deg}(\mathsf a)*\text{deg}(\mathsf b)}\ \mathsf b\otimes \mathsf a.
\end{equation}
See eq.~\eqref{eq:r-to-R} to map $r$ to a matrix $R^{IJ}$, with $I,J = 1, \ldots , \mathrm{dim}\mathfrak{g}$.

As already noted in~\cite{Borsato:2016ose}, using the above graded commutation relations, one can straightforwardly check the unimodularity condition \eqref{eq:uni-cond} (which, we recall, gives rise to supergravity backgrounds) for a generic $N$-extended Jordanian $R$-matrix, obtaining
\begin{equation}
R^{IJ}[[T_I,T_J]]=2(-1-N_0+N_1)\e,
\end{equation}
where $N_0, N_1$ are respectively the numbers of even and odd extra pairs of generators, so that $N_0+N_1=N$. Unimodularity  \eqref{eq:uni-cond} then implies $N_1=N_0+1$. 
Starting from a Jordanian $R$-matrix of the form $r=\h\wedge \e$, a minimal extension that makes it unimodular  will therefore be
\begin{equation}
r=\h\wedge \e- \e_+\wedge \e_-,
\end{equation}
with both $\e_\pm$ of odd grading (i.e.~$\text{deg}(\e_\pm)=1$, so that $N_0=0,N_1=1$)  and 
\begin{equation}
[\e_{\pm },\e]=0,\qquad
[\h,\e_\pm]=(\tfrac12\pm \xi)\e_\pm ,\qquad
\{\e_+,\e_-\}= \e ,
\end{equation}
with $\xi\in \mathbb C$.
In view of later calculations, we  find it convenient to redefine these odd elements as
\begin{equation}
    \Q_1=\frac{1}{\sqrt{2}}(\e_+-i\, \e_-),\qquad\qquad
    \Q_2=\frac{1}{\sqrt{2}}(i\, \e_+-\e_-),
\end{equation}
so that the (anti)commutation relations become
\begin{equation}\label{eq:Q-rel-N0-0}
[\Q_\is,\e]=0,\qquad
[\h,\Q_\is]=\tfrac12 \Q_\is -\varepsilon_{\is\js}\check\xi\,  \Q_\js,\qquad
\{\Q_\is,\Q_\js\}= -i\delta_{\is \js} \e,
\end{equation}
with $\is,\js=1,2$, $\check\xi= i \xi$ and the antisymmetric tensor $\varepsilon_{12}=-\varepsilon_{21}=1$. In this basis the $R$-matrix reads 
\begin{equation} \label{eq:rmatrix-rank2-uniext}
    r=\h\wedge\e-\frac{i}{2}(\Q_1\wedge\Q_1+\Q_2\wedge\Q_2).
\end{equation}
In the rest of the paper we will construct bosonic (i.e.~$N_1=0$) extended solutions with $N_0=0$ (i.e.~the standard rank-2 case constructed above), with $N_0=1$ (i.e.~rank-4 bosonic $R$-matrices) and with  $N_0=2$ (i.e.~rank-6 bosonic $R$-matrices), and we will find that $N_0>2$ is not possible. We will show that some $R$-matrices with $N_0=0$ and all of those that we construct with $N_0=1,2$ admit also  unimodular extensions (with $N_1=N_0+1$). In the generic case, we prefer to rewrite~\eqref{eq:ext-rel-Tol} in terms of $\e_{\pm \as}$ for the even pairs in the extension with  labels $\as,\bs=1,\ldots,N_0$, and $\Q_\is^\Is$ with $\is=1,2$ for the odd pairs in the extension with labels $\Is,\Js=1,\ldots,N_1$.
The relations  then read 
\begin{equation}\label{eq:ext-rel}
\begin{aligned}
&[\e_{\pm \as},\e]=0,\qquad &&[\h,\e_{\pm \as}]=(\tfrac12 \pm \hat\xi_\as )\e_{\pm \as},\quad 
&&[\e_{+ \as},\e_{- \bs}]=\delta_{\as \bs}\e,\quad 
&&[\e_{\pm \as},\e_{\pm \bs}]=0,\\
&[\Q_\is^\Is,\e]=0,\quad
&&[\h,\Q_\is^\Is]=\tfrac12 \Q_\is^\Is -\varepsilon_{\is\js}\check\xi^\Is\,  \Q_\js^\Is,\quad
&&\{\Q_\is^\Is,\Q_\js^\Js\}= -i\delta^{\Is\Js}\delta_{\is\js} \e,\quad 
&&[\Q^\Is_\is,\e_{\pm \as}]=0,
\end{aligned}
\end{equation}
and the $r$-matrix will be
\begin{equation}\label{eq:ext-r}
    r=\h\wedge\e+\sum_{\as=1}^{N_0} \e_{+ \as}\wedge \e_{-\as} -\frac{i}{2}\sum_{\Is=1}^{N_1} (\Q_1^\Is\wedge\Q_1^\Is+\Q_2^\Is\wedge\Q_2^\Is).
\end{equation}
In order to respect the reality conditions of $\alg g$ and, in our case,  to generate real deformations of the $AdS_5\times S^5$ background, we need to restrict the free parameters of the above relations to be real, $\hat\xi_\as,\check\xi^\Is\in \mathbb R$. We refer to appendix~\ref{app:conv} for more details.

\section{Classification of the non-extended solutions in $\alg{so}(2,4)$} \label{sec:rank2}

Our first task is to obtain the classification of all rank-2 (i.e.~non-extended with $N_0=N_1=0$) Jordanian solutions of the conformal algebra\footnote{The algebra spanned by $\h,\e$ is non-compact and thus neither $\h$ nor $\e$ can be an element of $\alg{so}(6)\subset\alg{psu}(2,2|4)$. Nevertheless, we can always shift $\h$ by en element of $\mathfrak{so}(6)$, and the classical Yang-Baxter equation will still be satisfied. This is true not only for the rank-2 solutions analysed in this section but also for the bosonic extensions of the next section.} $\alg{so}(2,4)\subset\alg{psu}(2,2|4)$. In other words, we are after all the \emph{inequivalent} choices of $\h$ and $\e$  among the  elements of $\alg{so}(2,4)$ that satisfy $[\h,\e]=\e$. Two choices $\{\h,\e\}$ and $\{\h',\e'\}$ are said to be equivalent if there exists an inner automorphism of $\alg{so}(2,4)$ that relates them, i.e.~if there exists an element $f\in SO(2,4)$ such that $\h '= f^{-1}\h f$ and $\e '= f^{-1}\e f$.

From the point of view of the classification of the deformations of $AdS_5\times S^5$,  modding out by  inner automorphisms is justified by how the $R$-matrix enters the action of the deformed $\sigma$-model, see \eqref{eq:hyb-ssssm}. In fact, $R$ appears in the linear operator $\mathcal O = 1 - \eta R_g \hat d:\g\to\g$, where $R_g=\AD_g^{-1} R \AD_g$.  In the undeformed model, multiplication of the supercoset representative $g\in G$ from the left by a constant element $f\in G$ corresponds to an isometry of the target-space background. In the presence of the deformation, under a left multiplication we have invariance of the $\sigma$-model action \eqref{eq:hyb-ssssm} up to a possible change of the $R$-matrix itself as $R\to \AD_f^{-1} R \AD_f$. Therefore, the group of isometries is reduced to the subgroup of $G$ that leaves $R$ invariant (i.e.~$R= \AD_f^{-1} R \AD_f$). Nevertheless,  when $f$ does not correspond to an isometry because it does not leave the $R$-matrix (and therefore the action)  invariant, multiplication by $f$ amounts  to just a  field redefinition of $g$, which is simply a different language to describe the same physics. 
Conversely, when two different deformations are generated by $R$ and $R'$ related as $R'=\AD_f^{-1} R \AD_f$, then it is enough to identify $g'=fg$ to conclude that the two deformations are physically equivalent. Notice that when $R'=\AD_f^{-1} R \AD_f$, then the relation between the Lie-algebra elements is implemented precisely by the adjoint $G$-action $T'_I= \AD_f^{-1}T_I$, a fact which justifies the definition of equivalence for the choices $\{\h,\e\}$ and $\{\h',\e'\}$.

According to the previous discussion, we must obtain all the embeddings of $\{\h,\e\}$ in $\alg{so}(2,4)$ modulo inner $SO(2,4)$ automorphisms. To do so, we will follow a strategy that was already used in~\cite{Borsato:2016ose} to classify all the inequivalent   rank-4 unimodular bosonic $R$-matrices of $\alg{so}(2,4)$. First, we notice that the 2-dimensional algebra generated by $\h$ and $\e$ is solvable. In~\cite{Patera:1973yn} it was proved that all solvable  subalgebras of $\alg{so}(2,4)$ must be subalgebras of one of the maximal solvable subalgebras\footnote{See the corollary at the end of section II.B of~\cite{Patera:1973yn}.} of $\alg{so}(2,4)$. The algebra $\alg{so}(2,4)$ has two non-abelian maximal solvable subalgebras\footnote{See the results of section III.D of~\cite{Patera:1973yn} for the group $SU(2,2)$ which is locally isomorphic to $SO(2,4)$, as well as~\cite{Patera1973b}. Notice that we are ignoring the maximal solvable subalgebra $\alg s_0$ because it is abelian. Moreover, following~\cite{Borsato:2016ose}, we swap the definition of $\alg s_1$ and  $\alg s_2$ compared to~\cite{Patera:1973yn}.} which, following~\cite{Borsato:2016ose}, we take as
\begin{equation}
\begin{aligned}
    \mathfrak{s}_1 &= \mathrm{span}(p_0,p_1,p_2,p_3 , J_{01}-J_{13},J_{02} - J_{23}, J_{03},J_{12},D) , \\
    \mathfrak{s}_2 &= \mathrm{span}(p_0+p_3 , p_1 , p_2, J_{01} - J_{13} , J_{02} - J_{23} , J_{12} , J_{03} - D, k_0 + k_3 + 2p_3 ) \ .
\end{aligned}
\end{equation}
Let us stress that also the identification of $\alg s_1$ and $\alg s_2$ is provided only up to inner automorphisms of $\alg{so}(2,4)$. That means that other choices are possible but they are all physically equivalent.

At this point we only need to find all the possible embeddings of the algebra generated by $\h$ and $\e$ in either $\alg s_1$ or $\alg s_2$, up to automorphisms generated by $SO(2,4)$. This task can be performed systematically by first identifying all possible embeddings of $\e$ up to inner automorphisms. The reason to single out this element first is that it is the only one that appears on the right-hand-side of the commutation relation. All such possible embeddings of $\e$ were already worked out in~\cite{Borsato:2016ose} and we recap them in Table \ref{tab:e-s1-s2}.

\begin{table}[h!]
    \centering
    \begin{tabular}{l|l}
       \multicolumn{1}{c|}{$\mathfrak{s}_1$}   & \multicolumn{1}{|c}{$\mathfrak{s}_2$}  \\
      \hline
    $(1)\ \  p_1 $     & $(1) \ \  p_1 $\\
    $(2)\ \  J_{02}-J_{23}$ &  $(2) \ \  p_0+p_3$ \\
   $(3)\ \ p_1 + J_{02}-J_{23}$  & $ (3) \ \ a\, p_1  + b\,  p_2 +J_{01} - J_{13} $\\
    $(4)\ \  p_0 $ &  \\
    $(5)\ \   p_3$  & \\
    $(6)\ \   p_0+p_3 $  & \\
    $(7)\ \  p_0-p_3+J_{01} - J_{13}$  & 
    \end{tabular}
    \caption{All the possible inequivalent embeddings of $\e$ in the two non-abelian maximal solvable subalgebras of $\alg{so}(2,4)$, up to inner $SO(2,4)$ automorphisms. In option $(3)$ of $\mathfrak{s}_2$, $a$ and $b$ are two real parameters.}
    \label{tab:e-s1-s2}
\end{table}

After fixing a choice for $\e$, it is a matter of imposing the commutation relation $[\h,\e]=\e$ to find $\h$ as well. After doing so, one must act again with inner  $SO(2,4)$ automorphisms  in order to remove as many free parameters as possible, and therefore identify all the \emph{inequivalent} embeddings of $\h$.\footnote{This is essentially the idea that was used in~\cite{Borsato:2016ose} to identify the possible embeddings of $\e$. Notice that in the case of $\e$ one does not include in the initial ansatz the generators that cannot appear on the right-hand-side of the commutation relations (i.e. $J_{03}$, $J_{12}$ and $D$ for $\mathfrak s_1$, and $J_{12}$, $J_{03}-D$ and $k_0+k_3+2p_3$ for $\mathfrak s_2$).}
In the following, when saying that we use an automorphism generated by $x\in \alg{so}(2,4)$ we mean that we implement the transformation $\h\to e^{-x}\h \,  e^x$ and $\e\to e^{-x}\e \,  e^x$. Given that the procedure assumes that the choice for $\e$ is fixed, we will only consider transformations that leave $\e$ invariant ($e^{-x}\e \,  e^x=\e$) or at most that they rescale it by an overall factor ($e^{-x}\e \,  e^x=c\, \e$) which can be reabsorbed by redefining the deformation parameter in the action.\footnote{This transformation, in fact, leaves $[\h,\e]=\e$ invariant. One could consider also $\e\to\e+c\, \h$, but it would generate new (non-physical) parameters rather than reabsorbing them.}  We also point out that, since $\mathfrak{s}_1$ and $\mathfrak{s}_2$ are \emph{maximal} solvable subalgebras of $\mathfrak{so}(2,4)$, the automorphisms we are after are in fact inner automorphisms of $\mathfrak{s}_1$ or $\mathfrak{s}_2$. In general, note that in $\h$ we can never remove contributions  from the last three generators of $\mathfrak{s}_1$ and $\mathfrak{s}_2$, because these generators never appear on the right-hand side of the commutation relations of these subalgebras of $\alg{so}(2,4)$.

We will describe in some detail the calculations for the first example, and we will summarise as briefly as possible the remaining ones.

\subsection{Embeddings in $\alg s_1$}
\subsubsection*{$(1) \ \ \e=p_1$}
In order to identify $\h$ one  starts from the parameterisation of a generic element of $\alg s_1$ and one takes for example $\h=\alpha^\mu p_\mu+\beta(J_{01}-J_{13})+\gamma(J_{02}-J_{23})+\delta J_{03}+\epsilon J_{12} +\lambda D$ with all the coefficients being generic real parameters. After imposing $[\h,\e]=\e$, one finds that some of the  parameters need to be fixed to special values. In particular we find $\h=D+\alpha^\mu p_\mu +\gamma(J_{02}-J_{23})+\delta J_{03}$. At this point we can act with inner $SO(2,4)$ automorphisms to further reduce the number of physical parameters. This is what we will explain in the following.

First of all, if $\delta\neq \pm 1$, then all contributions with $p_\mu $ in $\h$ can be removed by acting with an automorphism generated by $x=c^\mu p_\mu $ with $c^\mu \in \mathbb{R}$.
In fact, after noticing that $e^{-x}\e \,  e^x=\e$, one finds that
\begin{equation}
\begin{aligned}
    e^{-x}\h \, e^x&=D+(c^0-\gamma c^2-\delta c^3+\alpha^0)p_0+(c^1+\alpha^1)p_1+(c^2+\alpha^2-\gamma c^0+\gamma c^3)p_2\\
    &+(c^3+\alpha^3-\gamma c^2-\delta c^0)p_3+\gamma (J_{02}-J_{23})+\delta J_{03},
\end{aligned}
\end{equation}
so that to remove the contribution of $p_1$ it is enough to set $c_1=-\alpha_1$. Removing the contributions with $p_0,p_2,p_3$ is more delicate: one needs to impose a system of 3 linear equations for the 3 unknowns $c^0,c^2,c^3$. The determinant of the matrix associated to this linear system is $\det =1-\delta^2$. Therefore, if we assume that $\delta\neq \pm 1$ we can remove all contributions with $p_\mu $ as anticipated above.
At this point, an automorphism generated by $x=c(J_{02}-J_{23})$ is able to remove also $J_{02}-J_{23}$, at the extra condition that $\delta\neq 0$. Notice that if $\delta=0$ and $\gamma\neq 0$, then we can still set it to $\pm 1$ by acting with $x=cJ_{03}$.

If $\delta=1$, instead, after removing  $p_1$ by $x=c^1p_1$, we can also remove $J_{02}-J_{23}$ by $x=c(J_{02}-J_{23})$ because now we are sure to satisfy the condition $\delta\neq 0$. A further action with $x = c^\mu p_\mu $ (with $c^1=0$) allows us to remove $p_2$ and $p_0-p_3$ from $\h$, but not necessarily $p_0+p_3$. If not 0, the coefficient in front of $p_0+p_3$ can be set to $\pm 1$ by $x=cD$.
If $\delta=-1$ then the reasoning is similar, but with the role of $p_0-p_3$ and $p_0+p_3$ interchanged.

To recap, we have the following 3 inequivalent possibilities:
\begin{equation}
\begin{aligned}
    &1.1.a:\ \h=D+\delta J_{03},\\
    &1.1.b: \ \h = D+\gamma(J_{02}-J_{23}),\text{ with } \gamma^2=1,\\
    &1.1.c: \ \h=D+\alpha(p_0+\delta p_3)+\delta J_{03},\text{ with } \alpha^2=\delta^2=1.
\end{aligned}
\end{equation}

\subsubsection*{$(2) \ \ \e=J_{02}-J_{23}$}
Generically we may have $\h=\alpha^+(p_0+p_3)+\alpha^1p_1+\beta(J_{01}-J_{13})+\gamma(J_{02}-J_{23})+\lambda D-J_{03}$. First, we can remove both $J_{01}-J_{13}$ and $J_{02}-J_{23}$ by $x=c^1(J_{01}-J_{13})+c^2(J_{02}-J_{23})$. If $\lambda\neq 0$ and $\lambda\neq -1$, then we can also remove $p_0+p_3$ and $p_1$ by $x=c^+(p_0+p_3)+c^1p_1$. If $\lambda =0$ we can remove $p_0+p_3$ but not necessarily $p_1$, and the coefficient of the latter (if not 0) can be set to $\pm 1$ by $x=cD$. Similarly, if $\lambda =-1$ then we can remove $p_1$ but not necessarily $p_0+p_3$, and the coefficient of the latter can also be set to $\pm 1$.

To recap, we have the following 3 inequivalent possibilities:
\begin{equation}
\begin{aligned}
    &1.2.a:\ \h=-J_{03}+\lambda D,\\
    &1.2.b: \ \h = -J_{03}+\alpha p_1,\text{ with } \alpha^2=1,\\
    &1.2.c: \ \h=-J_{03}-D+\alpha(p_0+ p_3),\text{ with } \alpha^2=1.
\end{aligned}
\end{equation}

\subsubsection*{$(3) \ \ \e=p_1+J_{02}-J_{23}$}
Generically we may have $\h=D-J_{03}+\alpha^+(p_0+p_3)+\alpha^1p_1+\beta(p_2+J_{01}-J_{13})+\gamma(J_{02}-J_{23})$. First, we can act with an automorphism generated by $x=c(p_2+J_{01}-J_{13})$ to remove the contribution proportional to $\beta$. Similarly, acting with $x=cp_1$ will remove $p_1$, with $x=c(J_{02}-J_{23})$ will remove $J_{02}-J_{23}$ and with $x=c(p_0+p_3)$ will remove $p_0+p_3$.

In this example, therefore, there is only one possibility, namely
\begin{equation}
    \h=D-J_{03}.
\end{equation}

\subsubsection*{$(4) \ \ \e=p_0$}
We must start from $\h=D+\epsilon J_{12}+\alpha^\mu p_\mu $. Acting with $x=c^\mu p_\mu $ we can always remove all the $p_\mu $. The contribution with $J_{12}$, instead, cannot be removed and we have only one possibility
\begin{equation}
    \h=D+\epsilon J_{12}.
\end{equation}

\subsubsection*{$(5) \ \ \e=p_3$}
As in the previous example, we have $\h=D+\epsilon J_{12}+\alpha^\mu p_\mu $ and all contributions with $p_\mu $ can be removed with $x=c^\mu p_\mu $. We therefore have only one possibility
\begin{equation}
    \h=D+\epsilon J_{12}.
\end{equation}

\subsubsection*{$(6) \ \ \e=p_0+p_3$}
In general we have $\h=\alpha^\mu p_\mu +\beta(J_{01}-J_{13})+\gamma(J_{02}-J_{23})+\delta J_{03}+\epsilon J_{12} +(\delta+1) D$. If at least one of the two parameters $\delta,\epsilon$ is non-vanishing (i.e.~if $\delta^2+\epsilon^2\neq 0$) then we can remove the contributions with $J_{01}-J_{13},J_{02}-J_{23}$ with $x=c^1(J_{01}-J_{13})+c^2(J_{02}-J_{23})$. If we further assume that $\delta\neq -\tfrac12$ and that $(\delta+1)^2+\epsilon^2\neq0$ then we can also remove all $p_\mu $ with $x=c^\mu p_\mu $.

If instead $\delta=-\tfrac12$ we can remove all $p_\mu$ except the combination $p_0-p_3$, whose coefficient may be set to $\pm1$ with $x=cD$.

Another scenario is $\delta=-1$ and $\epsilon=0$. We can remove $p_0,p_3$ but not $p_1,p_2$ with $x=c^\mu p_\mu $. Nevertheless, if both $p_1,p_2$ are present, we can remove one of them by $x=cJ_{12}$, and set the coefficient of the remaining one to $\pm1$ by $x=cD$.

Finally, if $\delta=0$ and $\epsilon=0$ then we can act with $x=cJ_{12}$ to remove, for example, $J_{02}-J_{23}$. After doing that, we can always remove all $p_\mu $ by $x=c^\mu p_\mu $. At this point, it is possible to rescale the coefficient of $J_{01}-J_{13}$ by $x=cJ_{03}$.

To summarise, we have in total 4 possibilities:
\begin{equation}\label{eq:h-1.6}
    \begin{aligned}
         &1.6.a:\ \h=(\delta+1)D+\delta J_{03}+\epsilon J_{12},\\
         & 1.6.b:\ \h=\tfrac12(D-J_{03})+\epsilon J_{12}+\alpha(p_0-p_3),\text{ with } \alpha^2=1,\\
         &1.6.c:\ \h=-J_{03}+\alpha p_1,\text{ with } \alpha^2=1,\\
         & 1.6.d:\ \h= D+\alpha(J_{01}-J_{13}),\text{ with } \alpha^2=1.
    \end{aligned}
\end{equation}

\subsubsection*{$(7) \ \ \e=p_0-p_3+J_{01}-J_{13}$}
We start from $\h=2D-J_{03}+\alpha^+(p_0+p_3)+\alpha^2p_2+\beta(J_{01}-J_{13})+\gamma(J_{02}-J_{23})$. We can remove $J_{01}-J_{13},J_{02}-J_{23}$ with $x=c^1(J_{01}-J_{13})+c^2(J_{02}-J_{23})$,  $p_2$ with $x=cp_2$ and  $p_0+p_3$ with $x=c(p_0+p_3)$. Therefore, we only have one possibility
\begin{equation}
    \h=2D-J_{03}.
\end{equation}

\subsection{Embeddings in $\alg s_2$}

\subsubsection*{$(1) \ \ \e=p_1$}
We start from the parameterisation of a generic element of $\alg s_2$, namely $\h=\alpha^+(p_0+p_3)+\alpha^1p_1+\alpha^2p_2+\beta(J_{01}-J_{13})+\gamma (J_{02}-J_{23})+\delta(J_{03}-D)+\epsilon J_{12}+\lambda (k_0+k_3+2p_3)$. After imposing the commutation relation $[\h,\e]=\e$ we find $\h=D-J_{03}+\alpha^+(p_0+p_3)+\alpha^1p_1+\alpha^2p_2+\gamma (J_{02}-J_{23})$. We can remove the contribution with $J_{02}-J_{23}$ by acting with an automorphism generated by $x=c(J_{02}-J_{23})$. After that, we can remove all $p_\mu $ by $x=c^+(p_0+p_3)+c^1p_1+c^2p_2$. Therefore, we simply have
\begin{equation}
    \h=D-J_{03}.
\end{equation}
Actually this solution is the same found in case (1) of the $\alg s_1$ embedding. This simply means that the algebra that we are considering is a subalgebra of both $\mathfrak s_1$ and $\mathfrak s_2$.

\subsubsection*{$(2) \ \ \e=p_0+p_3$}
We have $\h=\tfrac12(D-J_{03})+\alpha^+(p_0+p_3)+\alpha^1p_1+\alpha^2p_2+\beta(J_{01}-J_{13})+\gamma (J_{02}-J_{23})+\epsilon J_{12}+\lambda (k_0+k_3+2p_3)$. We can remove the contributions with $J_{01}-J_{13},J_{02}-J_{23}$ by $x=c^1(J_{01}-J_{13})+c^2(J_{02}-J_{23})$, and similarly the contributions of $p_0+p_3,p_1,p_2$. Therefore we have
\begin{equation}\label{eq:h-s2-2}
    \h=\tfrac12(D-J_{03})+\epsilon J_{12}+\lambda (k_0+k_3+2p_3).
\end{equation}

\subsubsection*{$(3) \ \ \e=ap_1+bp_2+J_{01}-J_{13}$}
Here we need to distinguish different cases. If $a^2-b^2+1\neq 0$ or if $ab\neq 0$ then we have $\h=D-J_{03}+\alpha^+(p_0+p_3)+(a\beta+b\gamma)p_1+\alpha^2p_2+\beta(J_{01}-J_{13})+\gamma (J_{02}-J_{23})$. Now acting with $x=c^+(p_0+p_3)+c^1p_1+c^2p_2+c_\beta(J_{01}-J_{13})+c_\gamma (J_{02}-J_{23})$ we can remove all free parameters and have just
\begin{equation}
    2.3.a:\ \h=D-J_{03}.
\end{equation}
On the other hand, if $a=0$ and $b=\pm 1$ (which is the only real solution of the system $a^2-b^2+1= 0$ and $ab= 0$)  then we have $\h=D-J_{03}+\alpha^+(p_0+p_3)\pm \gamma p_1+\alpha^2p_2+\beta(J_{01}-J_{13})+\gamma (J_{02}-J_{23})\mp 2\lambda J_{12}+\lambda (k_0+k_3+2p_3)$. At this point we can act with $x=c^+(p_0+p_3)+c^1 p_1+c^2p_2+c_\beta(J_{01}-J_{13})+c_\gamma (J_{02}-J_{23})$ to remove several parameters and be left with only
\begin{equation}
   2.3.b:\   \h=D-J_{03}\mp 2\lambda J_{12}+\lambda (k_0+k_3+2p_3).
\end{equation}

\section{Classification of the (bosonic) extended solutions  in $\alg{so}(2,4)$} \label{sec:bos-ext}

The reasoning followed to classify the rank-2 solutions in $\alg{so}(2,4)$ can be applied also to find extended (i.e.~higher rank) bosonic Jordanian $R$-matrices. In fact, according to the commutation relations given in~\eqref{eq:ext-rel-Tol} or~\eqref{eq:ext-rel}, the $N$-extended algebra is also solvable, and if we want it to be a subalgebra of $\alg{so}(2,4)$ it must again be a subalgebra of either $\alg s_1$ or $\alg s_2$. Here we use this observation to classify the extended (higher-rank) Jordanian $R$-matrices. The strategy is to start from the classification of the rank-2 solutions of the previous section and construct elements $\e_{\pm \as}$ that satisfy~\eqref{eq:ext-rel} with a given $\h$ and $\e$ up to inner $\mathfrak{so}(2,4)$ automorphisms that now leave \textit{both} $\h$ and $\e$ invariant.

It turns out that for most choices of $\e\in\alg s_i$ it is not possible to identify two elements $\e_+,\e_-\in \alg s_i$ that commute with $\e$ and that satisfy $[\e_+,\e_-]=\e$. We therefore conclude that in those cases it is not possible to construct bosonic extended solutions. We find that it is possible to construct such solutions only when $\e=p_0+p_3$. This option shows up both in $\alg s_1$ and $\alg s_2$, and because it is a subset of the rank-2 solutions admitting a unimodular extension of Table~\ref{tab:rank-2-un}, we will refer to the names used in that table. Table~\ref{tab:rank-2-un} also summarises the inner $\mathfrak{so}(2,4)$ automorphisms that we can exploit.  When constructing the extended solutions we will refer to the names used in the summary of results of section~\ref{sec:summary}.

\subsection{Bosonic extensions in $\alg s_1$}

\subsubsection*{$N_0=1$}
Let us first try to construct an extension with $N_0=1$. We start from
\begin{equation}
    \e_\pm=\alpha_\pm^\mu p_\mu +\beta_\pm(J_{01}-J_{13})+\gamma_\pm(J_{02}-J_{23})+\delta_\pm(D+J_{03})+\epsilon_\pm J_{12},
\end{equation}
since these commute with $\e=p_0+p_3$. After imposing  by brute force the  relations in~\eqref{eq:ext-rel} with $\h$ given in $R_1$ of Table~\ref{tab:rank-2-un} (i.e.~case $1.6.a$ of~\eqref{eq:h-1.6}), we find that for $\hat \xi$ generic we can have the following solution\footnote{Notice that~\eqref{eq:ext-rel} is  symmetric under the transformation $\hat\xi_\as \to -\hat\xi_\as$ and $\e_{\pm \as}\to\pm \e_{\mp \as}$, so that in principle we also have the solution with 
$ \delta=-\hat\xi-\tfrac12,\ \epsilon=0$ and $\e_+=\bar\e_- ,\ \e_-=-\bar\e_+ $.
 We do not consider these as independent solutions because they only amount to a redefinition of the basis of the algebra. In fact, the $R$-matrix is unchanged under the transformation $\e_{\pm \as}\to\pm \e_{\mp \as}$. \label{f:symm-on-xi}}
\begin{equation}
\begin{aligned}
 &\delta=\hat\xi-\tfrac12,\quad\epsilon=0,\qquad (\implies \h=(1+\delta)D+\delta J_{03} ),\\
&   \e_+=\bar\e_+\equiv a\, p_1+b\,  p_2 ,\qquad
   \e_-=\bar\e_-\equiv c\,  (J_{01}-J_{13})+d\, (J_{02}-J_{23}) ,
   \end{aligned}    
\end{equation}
where the free (real) parameters are constrained to satisfy
\begin{equation}\label{eq:constr-ext-1}
    a\, c+b\, d=1.
\end{equation}
We may actually act with an automorphism generated by a $J_{12}$ rotation and remove the dependence on one of the above parameters. Without loss of generality\footnote{Because of the condition~\eqref{eq:constr-ext-1} we cannot have at the same time $c=0$ and $d=0$.} we may set  for example $c=0$, so that the quadratic condition reduces to $d=b^{-1}$. At this point, we can act with an automorphism generated by $D+J_{03}$ which allows us to set $b=\pm 1$. Taking into account that the transformation $\e_\pm\to c^{\pm 1}\e_\pm$ for any $c\in\mathbb{R}$ is an automorphism of the algebra and leaves the $R$-matrix invariant, we can effectively set $b=1$.  This is the solution $R_7$ in Table~\ref{tab:rank-4}.

When $\hat \xi$ takes some special values, the solutions for $\e_\pm$ can be slightly more generic. First we find\footnote{As already remarked, the relations in~\eqref{eq:ext-rel} are symmetric under $\hat\xi_\as\to -\hat\xi_\as$ and $\e_{\pm \as}\to\pm \e_{\mp \as}$. Therefore we have also solutions obtained  by this transformation, namely with  $\hat\xi=-\tfrac12$ and $\hat\xi=-\tfrac16$. These are not independent solutions, see footnote \ref{f:symm-on-xi}. In principle, we also find additional solutions with either $\e_+$ or $\e_-$ containing a contribution proportional to $p_0+p_3$, but this can be shifted away by an automorphism, so that the solution reduces to a special case of $R_7$.}
\begin{equation}
    \begin{aligned}
        & \hat\xi=\tfrac16,\quad \delta=-\frac13,\quad \epsilon=0,\quad (\implies \ \h=\tfrac23 D-\tfrac13 J_{03}),\\
        & \e_-=f(p_0-p_3)+\bar\e_- , \qquad  \e_+=\bar\e_+,
 \end{aligned}
\end{equation}
still subject to the constraint~\eqref{eq:constr-ext-1} and with the parameter $f\in \mathbb{R}$ free. As done previously, we can set $c=0$ by $J_{12}$, then $b=\pm 1$ by  $D+J_{03}$, and then send $\e_\pm\to b^{\pm 1}\e_\pm$ yielding $R_8$ in Table~\ref{tab:rank-4}.

Second, we   also have the option
\begin{equation}
    \begin{aligned}
      & \hat\xi=0,\qquad\delta=-\frac12,\qquad \epsilon=0,\quad (\implies\ \h=\tfrac12 (D-J_{03})),\\
       &\e_+=a_+p_1+b_+p_2+c_+(J_{01}-J_{13})+d_+(J_{02}-J_{23}) ,\\
    &\e_-=a_-p_1+b_-p_2+c_-(J_{01}-J_{13})+d_-(J_{02}-J_{23}),
    \end{aligned}
\end{equation}
with the more general constraint
\begin{equation}
    a_+c_-+b_+d_--a_-c_+-b_-d_+=1.
\end{equation}
For this constraint to admit a solution, notice  that either $\e_+$ or $\e_-$ must have a $p_\mu $ with non-vanishing coefficient, and the other must have $J_{0i}-J_{i3}$ with non vanishing coefficient. For definiteness, let us assume that $a_+\neq 0$ and $c_-\neq 0$. The other options are obtained by applying the symmetries $\e_\pm\to \pm \e_\pm$ or $1\leftrightarrow 2$ of the spatial indices 1,2, that give rise to equivalent solutions. 
As we can see in Table \ref{tab:rank-2-un}, in this case $\h$ and $\e$ are invariant not only under the action of $D+J_{03}$ and  $J_{12}$ but also $p_0-p_3$ and $k_0+k_3$. It should be possible to  use the last 3 of these generators to set $c_+=b_+=b_-=0$, so that the quadratic constraint reduces to $c_-=a_+^{-1}$. After that,  using $D+J_{03}$ one could set $a_+=\pm 1$. In other words, we believe it is possible to reduce this case to
\begin{equation}
    \begin{aligned}
      & \hat\xi=0,\qquad\delta=-\frac12,\qquad \epsilon=0,\quad (\implies\ \h=\tfrac12 (D-J_{03})),\\
       &\e_+=\alpha p_1+d_+(J_{02}-J_{23}) ,\\
    &\e_-=a_-p_1+\alpha (J_{01}-J_{13})+d_-(J_{02}-J_{23}),\qquad \text{with }\alpha^2=1.
    \end{aligned}
\end{equation}
However, it is quite subtle to make sure that setting $c_+=b_+=b_-=0$ is  possible for all values of the initial parameters $a_\pm,b_\pm,c_\pm,d_\pm$. In fact, one may  worry about possible singularities for special values of these parameters. The actions of the generators $J_{12},p_0-p_3,k_0+k_3$ mix non-trivially, and this makes the analysis more complicated. For this reason,  the solution $R_9$ in Table~\ref{tab:rank-4} is presented without the maximal simplification by automorphisms. We simply set $c_+=0$ by means of $k_0+k_3$, then set $a_-=0$ by means of $p_0-p_3$, then set $a_+=\pm 1$ thanks to $D+J_{03}$ and finally send $\e_\pm\to a_+^{\pm 1}\e_\pm$.

The calculations to identify the possible $N_0=1$ extensions of $R_2$ (i.e.~the case $1.6.b$ in~\eqref{eq:h-1.6}) are similar to the above ones (when setting $\hat\xi=0,\delta=-\tfrac12$). However, no solution is possible because $p_0-p_3$ does not commute with $J_{01}-J_{13}$ and $J_{02}-J_{23}$, and therefore $R_2$ does not admit a bosonic extension. Also the calculation for $R_3$ (i.e.~case $1.6.c$) is  similar to the ones above (now setting $\delta=-1$) and we can obtain the solution $R_{10}$ of Table~\ref{tab:rank-4} by borrowing the extension $R_7$ (\emph{before} removing any parameter by conjugation) at the condition that we further set $c=0$ in $\bar\e_-$ (because of the extra contribution of $p_1$ in $\h$). 
Similarly, $R_4$ (i.e.~case $1.6.d$) admits an extension as in $R_{7}$ at the condition of further setting $a=0$ in $\bar\e_+$, giving rise to the solution $R_{11}$. Note that, both in the case of $R_{10}$ and $R_{11}$,  $J_{12}$ and $D+J_{03}$ do not commute with $\h$. Therefore, one can not use these   to remove and rescale parameters.

\subsubsection*{$N_0=2$}
Using the above results, we can check that it is  possible to construct also extended bosonic Jordanian solutions with $N_0=2$. In particular, for generic $\hat\xi$ we can start from $R_7$ and construct the solution\footnote{In principle each pair $\{\e_{+\as},\e_{-\as}\}$ may come with its own coefficient $\hat\xi_\as$, but $\delta$ is already constrained to be $\delta=\pm \hat\xi-\tfrac12$ which implies that $\hat\xi_1=\hat\xi_2=\hat\xi$. We could in fact combine different solutions related by $\hat\xi\to-\hat\xi$ but they would be equivalent to those that we write here.} 
\begin{equation}
    \begin{aligned}
          R_{15}: \   &\delta=\hat\xi-\tfrac12,\quad\epsilon=0,\qquad (\implies \h=(\hat\xi+\tfrac12)D+(\hat\xi-\tfrac12)J_{03}),\\
   &\e_{+\as}=\bar \e_{+\as}\equiv a_\as p_1+b_\as p_2 ,\qquad
   \e_{-\as}= \bar \e_{-\as}\equiv c_\as (J_{01}-J_{13})+d_\as(J_{02}-J_{23}) ,\\
    \end{aligned}
\end{equation}
with the simultaneous conditions
\begin{equation} \label{eq:16a-rank6-constraints}
      a_2c_1+b_2d_1=0,\qquad 
    a_1c_2+b_1d_2=0,\qquad
      a_\as c_\as +b_\as d_\as =1,\quad \as =1,2.
\end{equation}
 Using the automorphism generated by a $J_{12}$ rotation we can set for example $c_1=0$ which then implies $b_2=0$ and $d_1=1/b_1,c_2=1/a_2,d_2=-a_1/(b_1a_2)$. Moreover, using $D+J_{03}$ we can set $b_1=\pm1$ and finally taking into account that also the transformation $\e_{\pm \as}\to c^{\pm 1}\e_{\pm \as}$ is an automorphism of the algebra (and leaves the $R$-matrix invariant) we can simply set $a_2=1$. This gives the solution $R_{13}$ of section~\ref{sec:summary}. 
 
 It is not possible, instead, to construct an extended solution from $R_{8}$ because $p_0-p_3$ does not commute with $J_{01}-J_{13}$ nor $J_{02}-J_{23}$. Therefore we cannot construct two linearly independent elements $\e_{-1},\e_{-2}$ unless they both have a vanishing coefficient in front of $p_0-p_3$ ($f_\as=0$), which then reduces to the previous solution. Finally, from $R_9$ we can construct the following extended solution
\begin{equation}
    \begin{aligned}
   R_{14}: \    & \hat\xi=0,\qquad\delta=-\frac12,\qquad \epsilon=0,\qquad (\implies \ \h=\tfrac12(D-J_{03})),\\
       &\e_{\pm \as}=a_{\pm \as}p_1+b_{\pm \as}p_2+c_{\pm \as}(J_{01}-J_{13})+d_{\pm \as}(J_{02}-J_{23}) ,
    \end{aligned}
\end{equation}
with the 6 constraints
\begin{equation}
\begin{aligned}
     &   a_{\pm 1}c_{\pm 2}+b_{\pm 1}d_{\pm 2}-b_{\pm 2}d_{\pm 1}-a_{\pm 2}c_{\pm 1}=0,\\
     &   a_{\pm 1}c_{\mp 2}+b_{\pm 1}d_{\mp 2}-b_{\mp 2}d_{\pm 1}-a_{\mp 2}c_{\pm 1}=0,\\
&    a_{+\as}c_{-\as}+b_{+\as}d_{-\as}-a_{-\as}c_{+\as}-b_{-\as}d_{+\as}=1,\quad \as=1,2.
    \end{aligned}
\end{equation}
These are all the options for extended solutions with $N_0=2$. It is not possible to construct them from $R_3$ and $R_4$ (or equivalently $R_{10}$ and $R_{11}$) because the extra condition of not having either $J_{01}-J_{13}$ or $p_1$ in $\e_{\pm \as}$ implies that there would not be enough linearly-independent vectors to construct the full solution.

For a similar reason, it is obviously not possible to construct extended solutions with $N_0>2$. There would not be enough linearly-independent vectors to construct the pairs $\{\e_{+\as},\e_{-\as}\}$.

\subsection{Bosonic extensions in $\alg s_2$}

\subsubsection*{$N_0=1$}

To understand whether we can construct a bosonic extension in this case, we proceed as before starting from 
\begin{equation}
    \e_\pm=\alpha^+_\pm(p_0+p_3)+\alpha^1_\pm p_1+\alpha^2_\pm p_2+\beta_\pm (J_{01}-J_{13})+\gamma_\pm  (J_{02}-J_{23})+\epsilon_\pm  J_{12}+\lambda_\pm  (k_0+k_3+2p_3),
\end{equation}
that commute with $\e$. We notice that this is a slight modification of the calculation done for the embedding in $\alg s_1$: first there is no combination $p_0-p_3$, and second there is an additional contribution of $k_0+k_3+2p_3$ both in $\e_\pm$ and in $\h$. In order to find a bosonic  extension to this solution that is new compared to what found in $\alg s_1$, we must therefore have $\lambda\neq 0$ in $\h$ or $\lambda_+\neq 0$ in $\e_+$ or $\lambda_-$ in $\e_-$ (or more than one of these possibilities simultaneously). We find that it is possible to construct extended solutions of this kind at $\hat\xi=0$  if we set $\lambda=\pm \frac{\epsilon}{2}$ in~\eqref{eq:h-s2-2} and if we take
\begin{equation}
\begin{aligned}
      & \e_+=a_+p_1+b_+p_2\mp b_+(J_{01}-J_{13})\pm a_+(J_{02}-J_{23}),\\
&  \e_-=a_-p_1+b_-p_2\mp b_-(J_{01}-J_{13})\pm a_-(J_{02}-J_{23}),
\end{aligned}
\end{equation}
with the condition
\begin{equation}
    a_+b_--a_-b_+=\mp \tfrac12,
\end{equation}
where the signs are correlated to the choice of sign in $\lambda=\pm\tfrac{\epsilon}{2}$. Note that to have a genuinely new extension (compared to the embedding in $\mathfrak s_1$) we must assume $\epsilon\neq 0$.  Although both $k_0+k_3-p_0+p_3$ and $J_{12}$ leave $\h$ and $\e$ invariant, we can remove only one parameter by their actions (because it turns out that there is only one non-trivial parameter when applying the two actions simultaneously) and we decide to set $b_+=0$. Then the constraint is solved just by $b_-=\mp \frac{1}{2a_+}$. Nevertheless,   we can always shift $\e_-$ by a quantity proportional to $\e_+$, given that at the level of the $R$-matrix the contribution of this extra shift drops out because of antisymmetry of $R$.\footnote{Notice that this is an automorphism of the algebra only if $\hat\xi=0$, but we could use this transformation even if $\hat\xi\neq 0$ because ultimately we are interested in the classification of the $R$-matrices rather than the algebras.} Therefore, the contribution proportional to $a_-$ produces no effect, and we can just set $a_-=0$. This is the solution $R_{12}$ in Table~\ref{tab:rank-4}.

In this case it is not possible to construct bosonic extended solutions with $N_0>1$. There are simply not enough parameters to satisfy all needed conditions.

\section{Classification of the unimodular solutions in $\alg{psu}(2,2|4)$} \label{sec:uni-ext}

In this section we address the question of whether it is possible to extend the bosonic Jordanian solutions constructed above to obtain unimodular solutions. As recalled in section~\ref{sec:jord}, we should add $N_1$ pairs of odd generators from $\alg{psu}(2,2|4)$ with $N_1=N_0+1$, where $N_0$ is the number of extra bosonic pairs of generators in the bosonic extension.

We will construct odd elements as linear combinations of the supercharges  of $\alg{psu}(2,2|4)$
\begin{equation}
    \Q^\Is_\is=q^\Is_\is\cdot Q +\overline q^\Is_\is\cdot \overline Q+s^\Is_\is\cdot S +\overline s^\Is_\is\cdot \overline S.
\end{equation}
Here we are using a simplifying notation where the dot product means that the indices of the supercharges are assumed to be in their canonical position $Q_{\alpha a},\overline Q^{\dot\alpha a},S_\alpha{}^a,\overline S^{\dot \alpha}{}_a$ and are contracted by complex coefficients $q,\overline q,s,\overline s$ (where we omit the obvious $\Is,\is$ indices for simplicity) with Lorentz and $\mathcal R$-symmetry indices in appropriate positions.\footnote{This means that $q\cdot Q=q^{\alpha a}Q_{\alpha a}$, $\overline q\cdot \overline Q=\overline q_{\dot\alpha a}\overline Q^{\dot\alpha a}$, $s\cdot S=s^\alpha{}_aS_\alpha{}^a$ and $\overline s\cdot\overline S=\overline s_{\dot \alpha}{}
^a\overline S^{\dot \alpha}{}_a$. As explained in appendix~\ref{app:psu224}, Lorentz indices $\alpha,\dot\alpha$ are raised and lowered with $\epsilon$ while the $\mathcal R$-indices $a$ are raised and lowered with the matrix $K$. Given that both matrices are antisymmetric, the contraction preserves the sign if both indices are changed of position, e.g.~$q\cdot Q=q^{\alpha a}Q_{\alpha a}=q_{\alpha a}Q^{\alpha a}$. Otherwise the overall sign changes.}
Importantly, we will demand that the odd elements $\Q^\Is_\is$ satisfy the standard reality conditions of $\alg{psu}(2,2|4)$, namely $(\Q^\Is_\is)^\dagger+H\Q^\Is_\is H^{-1}=0$, see appendix~\ref{app:psu224} for more details. Given that the supercharges that we are using for the superalgebra basis are not real and satisfy~\eqref{eq:reality-QS} instead, this implies that  the complex  coefficients $q,\overline q,s,\overline s$ must be such that 
\begin{equation}
    (q^{\alpha a})^*=-\epsilon^{\alpha\beta}\delta^{ab}\overline q_{\beta b},\qquad\qquad
    (s^\alpha{}_a)^*=-\epsilon^{\alpha\beta}\delta_{ab}\overline s_\beta{}^b,
\end{equation}
where the star is the complex conjugation.\footnote{Importantly, here the $a,b$ indices are raised and lowered with the Kronecker $\delta$ rather than the matrix $K$.}
In total we therefore have 32 real coefficients for each $\Q^\Is_\is$. Alternatively, we may define the real supercharges\footnote{This definition breaks Lorentz and $\mathcal R$-symmetry covariance but is useful for practical calculations in our case.}
\begin{equation}
    \mathcal Q_\pm^{\alpha a}\equiv \overline Q^{\alpha a}\pm \epsilon^{\alpha\beta}\delta^{ab}Q_{\beta b},\qquad\qquad 
    \mathcal S_\pm^\alpha{}_a\equiv  \overline S^\alpha{}_a\pm \epsilon^{\alpha\beta}\delta_{ab}S_\beta{}^b,
\end{equation}
and rewrite 
\begin{equation} \label{eq:def-Q-real-coeffs}
    \Q^\Is_\is=q^{\Is,+}_{\is}\cdot \mathcal Q_+ +i\, q^{\Is,-}_{\is}\cdot \mathcal Q_-+s^{\Is,+}_{\is}\cdot \mathcal S_+ +i\, s^{\Is,-}_{\is}\cdot \mathcal S_-,
\end{equation}
where now the coefficients $q^\pm$ and $s^\pm$ are simply real. We will present our derivation in this basis, because it makes it straightforward to check when the reality conditions present an obstruction to the unimodular extension.

\subsection{Unimodular extension of rank-2 solutions} \label{sec:rank2-uniext}

One can go through Tables~\ref{tab:rank-2-un} and~\ref{tab:rank-2-no-un}, which contain the summary of all  the rank-2 bosonic solutions, and check when it is possible to construct  unimodular extensions. In order to do that, we must construct two odd elements
\begin{equation}
    \Q_\is=q^{+}_{\is}\cdot \mathcal Q_+ +i\, q^{-}_{\is}\cdot \mathcal Q_-+s^{+}_{\is}\cdot \mathcal S_+ +i\, s^{-}_{\is}\cdot \mathcal S_-,
\end{equation}
with $\is=1,2$ that satisfy the relations in~\eqref{eq:Q-rel-N0-0} for a given $\h$ and $\e$. If this is possible, we will denote the unimodular extension of $R$ by $\bar{R}$.

\subsubsection*{Obstructions to the unimodularity extension}
Let us present explicitly the calculations for  one case that does not admit a unimodular extension, namely $\h=D+a J_{03}$ and $\e=p_1$. Notice that the argument for the obstruction presented in~\cite{vanTongeren:2019dlq} follows essentially the same logic as the explicit calculations that we do here. In general, the best strategy is to first impose the conditions in~\eqref{eq:Q-rel-N0-0} that are linear in $\Q_\is$. Starting with $[\Q_\is,\e]=0$ one immediately finds that  this case  implies $s^\pm=0$, so that $\Q_\is$ can be only linear combinations of the supercharges $\mathcal Q_\pm$. After imposing also $[\h,\Q_\is]=\tfrac12 \Q_\is-\epsilon_{\is\js}\, \xi\, \Q_\js$, one finds that the only option that (potentially) does not break the reality of the free coefficients $a,\xi,q^\pm$ is setting just $\xi=a=0$. At this point we turn to the relation quadratic in the odd elements, namely $\{\Q_\is,\Q_\js\}=-i\delta_{\is\js}\e$. This is where the real basis  for the supercharges turns out to be useful. Omitting the $\is$ index now, we find that, in  order to satisfy $\{\Q,\Q\}=-i\e$, the following equations must hold  
\begin{equation}
\begin{aligned}
     &\sum_{a=1}^4(q^+_{\alpha a})^2+\sum_{a=1}^4(q^-_{\alpha a})^2=0,\qquad \alpha=1,2,\\
     &\sum_{a=1}^4(q^+_{1a}q^+_{2a}+q^-_{1a}q^-_{2a})=-\tfrac14,\qquad 
     \sum_{a=1}^4(q^+_{1a}q^-_{2a}-q^-_{1a}q^+_{2a})=0 .
\end{aligned}
\end{equation}
As the coefficients are real,  the  two equations in the first line are solved only by $q^\pm_{\is,\alpha a}=0$. This solution is not compatible with the first equation of the second line, and in fact makes $\Q_\is$ vanish completely. Therefore for $\h=D+a J_{03}$ and $\e=p_1$ it is not possible to construct a unimodular extension.

The calculations are similar for all other cases of Table~\ref{tab:rank-2-no-un}, and we will not present the details for all of them. The obstructions   originate from reality conditions either from the linear commutation relations with $\h$ or from the quadratic anticommutation relations. This remains to be the case also in the presence of possible $\mathfrak{so}(6)$ shifts of $\h$. As indicated in Table~\ref{tab:rank-2-un}, however,  it is possible to construct unimodular extensions when $\e=p_0$ or $\e=p_0+p_3$, as we will now show.

\subsubsection*{Allowed unimodular extensions}
Let us start from those with $\e=p_0+p_3$, namely from $R_1$ to $R_5$ of Table \ref{tab:rank-2-un} included.
With this choice,  imposing $[\Q_\is,\e]=0$ sets $s^\pm_{\is,1a}=0$ and thus kills a total of 16 parameters. To proceed we need to specify the choice for $\h$ in each $R_i$.

\subsubsection*{$R_1$ } 
When taking $\h=(1+b)D+b J_{03}+a J_{12}$  and imposing $[\h,\Q_\is]=\tfrac12 \Q_\is-\epsilon_{\is\js}\, \check\xi\, \Q_\js$ we find that we  must set $\check{\xi}=\frac{a}{2}$ and that we have six branches for the solutions, depending on whether the parameters $a$, $b$ and/or $b+1$ are generic or vanishing.\footnote{As in the bosonic case, the relations in~\eqref{eq:ext-rel} are symmetric under $\check\xi\to-\check\xi$ and $\Q_1\leftrightarrow \Q_2$. That means that we also have solutions with $\check\xi=-a/2$. We do not write these explicitly because they give rise to the same $R$-matrix and deformation as the above solution. \label{f:symm-on-xi-uni}}  Nevertheless, after the quadratic commutation relations the six branches collapse to only two branches, $a$ is generic or $a=0$. 

\noindent $\bullet$ $a=0$ 

If $b\neq 0,-1$ we find that we must set
\begin{equation}\label{eq:unext-16a-1.1}
    s^\pm_{\is,\alpha a}=0,\qquad 
    q^\pm_{\is,1a}=0.
\end{equation}
The non-trivial coefficients $q^\pm_{\is,2a}$  will be further constrained by the condition $\{\Q_\is,\Q_\js\}=-i\delta_{\is\js}\e$. In particular we find
\begin{equation}\label{eq:unext-16a-1.2}
    \sum_{a}[(q^+_{\is,2a})^2+(q^-_{\is,2a})^2]=\tfrac12,\quad \text{for } \is=1,2,\qquad\  
    \sum_a (q^-_{1,2a}q^-_{2,2a}+q^+_{1,2a}q^+_{2,2a})=0 .
\end{equation}
If $b=0$ we find  less-restrictive conditions from the linear commutation relations with $\e$ and $\h$, namely  $s^\pm_{\is,\alpha a}=0$, but  real solutions of the anti-commutation relations reduce this case back to \eqref{eq:unext-16a-1.1} and \eqref{eq:unext-16a-1.2}. Similarly, if  $b=-1$, the linear commutation relations set only $s^\pm_{\is,1 a}=q^\pm_{\is,1a}=0$ but reality of the anti-commutation relations reduce this case back to \eqref{eq:unext-16a-1.1} and \eqref{eq:unext-16a-1.2}. We will denote this unimodular extension by $\bar{R}_1$.

\noindent $\bullet$ $a$ generic 

If $b\neq 0,-1$  we find from the linear commutation relations with $\e$ and $\h$ that we must set
\begin{equation}\label{eq:unext-16a-2.1}
    s^\pm_{\is,\alpha a}=0,\qquad 
    q^\pm_{\is,1a}=0,\qquad 
    q^+_{\is,2a}= \epsilon_{\is\js}q^-_{\js,2a}.
\end{equation}
The remaining quadratic conditions $\{\Q_\is,\Q_\js\}=-i\delta_{\is\js}\e$ can then be written as equations for the coefficients $q^-_{\is,2a}$ only, reading
\begin{equation}\label{eq:unext-16a-2.2}
    \sum_{\is,a}(q^-_{\is,2a})^2=\tfrac12.
\end{equation}
If $b=0$ we find again  a less restrictive solution for the linear commutation relations, namely
    $s^\pm_{\is,\alpha a}=0,$  
    $q^+_{2,1a}=q^-_{1,1a}=0,$ $  q^+_{1,1a}=-q^-_{2,1a}$, 
    $q^+_{\is,2a}= \epsilon_{\is\js}q^-_{\js,2a}$.
As before, to solve the quadratic equations $\{\Q_\is,\Q_\js\}=-i\delta_{\is\js}\e$ over the real numbers, however, we must set $q^+_{1,1a}=0$ and we reduce back to the previous case \eqref{eq:unext-16a-2.1} and \eqref{eq:unext-16a-2.2}. Similarly, if  $b=-1$,  solving the linear and quadratic equations over the reals reduces the conditions to \eqref{eq:unext-16a-2.1} and \eqref{eq:unext-16a-2.2}. We will denote this unimodular extension by $\bar{R}_{1'}$.

\vspace{.2cm}

\noindent In summary, we only have to distinguish  the $a=0$ case, enforcing \eqref{eq:unext-16a-1.1} and \eqref{eq:unext-16a-1.2}, and the case in which $a$ is generic, enforcing \eqref{eq:unext-16a-2.1} and \eqref{eq:unext-16a-2.2}.  Both cases admit solutions over the real numbers. Note that \eqref{eq:unext-16a-2.1} and \eqref{eq:unext-16a-2.2} will solve \eqref{eq:unext-16a-1.1} and \eqref{eq:unext-16a-1.2}, but not the other way around, and therefore  the solutions $\bar{R}_{1'}$ reduce to a \emph{special} solution of the $\bar{R}_{1}$ when taking the limit $a\to 0$.

\subsubsection*{$R_2$} 
It turns out that the calculations for the case $\h=\tfrac12(D-J_{03})+a J_{12}+\alpha(p_0-p_3)$ with $\alpha^2=1$ are completely analogous to the previous one. In fact, the parameter $b$ of $R_1$ plays no role in the classification of the solutions, and the extra contribution with $p_0-p_3$ is harmless because the supercharges $Q,\overline Q$ commute with the momenta $p_\mu$. If $a=0$, the solutions are then given by~\eqref{eq:unext-16a-1.1} and~\eqref{eq:unext-16a-1.2}, whose corresponding $R$-matrix we denote by $\bar{R}_2$, or if $a$ is generic they are given by~\eqref{eq:unext-16a-2.1} and~\eqref{eq:unext-16a-2.2}, denoted by $\bar{R}_{2'}$.

\subsubsection*{$R_3$} 
The calculations for $\h=-J_{03}+\alpha p_1$ with $\alpha^2=1$ are analogous to the case $R_1$ with the condition $a=0$, so that the solutions are given by~\eqref{eq:unext-16a-1.1} and~\eqref{eq:unext-16a-1.2}, whose $R$-matrix will be denoted by $\bar{R}_3$.

\subsubsection*{$R_4$} 
The calculations for $\h= D+\alpha(J_{01}-J_{13})$ with $\alpha^2=1$ are also analogous to the case $R_1$ with the condition $a=0$. In fact, the element $J_{01}-J_{13}$ has non-vanishing commutators with supercharges with $\alpha=1$, but these contributions are already set to zero by other conditions. Therefore, the solutions are given by~\eqref{eq:unext-16a-1.1} and~\eqref{eq:unext-16a-1.2}, whose $R$-matrix will be denoted by $\bar{R}_4$.

\subsubsection*{$R_5$}
Here we have $\h=\tfrac12(D-J_{03})+a J_{12}+b(k_0+k_3+2p_3)$ and $\e=p_0+p_3$, which compared to $R_1$ is a genuinely new case only if $b\neq 0$.
The results  are in fact identical to  $R_1$, and therefore if $a=0$ they are given by~\eqref{eq:unext-16a-1.1} and~\eqref{eq:unext-16a-1.2}, whose $R$-matrix we denote by $\bar{R}_5$,  and if $a$ is generic they are given by~\eqref{eq:unext-16a-2.1} and~\eqref{eq:unext-16a-2.2}, with $R$-matrix $\bar{R}_{5'}$.

\subsubsection*{$R_6$}
Let us now consider the choice $\h=D+aJ_{12}$ and $\e=p_0$. Imposing $[\Q_\is,\e]=0$ we  find  that all  $s^\pm=0$. After imposing  $[\h,\Q_\is]=\tfrac12 \Q_\is-\epsilon_{\is\js}\, \check\xi\, \Q_\js$ we conclude that we must impose $\check\xi=\frac{ a}{2}$ and that we have two branches:

\noindent $\bullet$ $a=0$

In this case, the  commutation relation for $[\h, \Q_\is]$ is automatically satisfied. The only constraints on the coefficients $q^\pm$ then follow from the quadratic relations $\{\Q_\is, \Q_\js\} = -i \delta_{ij}\e$. They read
\begin{equation} \label{eq:quadr6a0}
\begin{alignedat}{2}
    &\sum_{a}[(q^-_{\is,\alpha a})^2 +(q^+_{\is,\alpha a})^2]=\tfrac14,\qquad &&\is=1,2,\qquad \alpha=1,2,\\
    &\sum_{a}( q^-_{\is,1a}q^-_{\is,2a}+ q^+_{\is,1a}q^+_{\is,2a})=0, &&\is=1,2,\\
    &\sum_a \epsilon^{\alpha\beta} q^+_{\is,\alpha a} q^-_{\is,\beta a}=0, \qquad &&\is=1,2,\\
     &\sum_{a}(q^-_{1,\alpha a}q^-_{2,\alpha a}+q^+_{1,\alpha a}q^+_{2,\alpha a})=0,\qquad &&\alpha=1,2,\\
     &\sum_{a}\sigma_1^{\alpha\beta}(q^-_{1,\alpha a}q^-_{2,\beta a}+q^+_{1,\alpha a}q^+_{2,\beta a})=0, \qquad && \\ &\sum_{a} \epsilon^{\alpha\beta}(q^+_{1,\alpha a}q^-_{2,\beta a}+q^+_{2,\alpha a}q^-_{1,\beta a})=0 ,\qquad  &&
    \end{alignedat}
\end{equation}
which admit solutions over the real numbers.
We will denote this unimodular extension by $\bar{R}_6$. 

\noindent $\bullet$ $a$ generic

In this case $[\h, \Q_\is]$ imposes
\begin{equation}
    q^+_{\is,\alpha a}= (-1)^\alpha \epsilon_{\is\js} q^-_{\js,\alpha a}.
\end{equation}
We are thus able to solve for all the $q^+$ coefficients, for example, so that $\{\Q_\is,\Q_\js\}=-i\delta_{\is\js}\e$ can be imposed  on the coefficients $q^-_{\is,\alpha a}$ only. They read
\begin{equation}\label{eq:quadp0p3}
    \begin{aligned}
     &\sum_{\is,a}(-1)^\is\, q^-_{\is,1a}q^-_{\is,2a}=0,\qquad\qquad 
     \sum_{a=1}^4(q^-_{1,1a}q^-_{2,2a}+q^-_{1,2a}q^-_{2,1a})=0,\\
     & \sum_{\is,a}(q^-_{\is,\alpha a})^2=\tfrac14,\qquad \alpha=1,2 ,
    \end{aligned}
\end{equation}
which admit solutions over the real numbers.  We will denote this unimodular extension by $\bar{R}_{6'}$. 

Note that imposing $q^+_{\is,\alpha a}= (-1)^\alpha \epsilon_{\is\js} q^-_{\js,\alpha a}$ on \eqref{eq:quadr6a0} these set of equations reduced to the set \eqref{eq:quadp0p3}. This means that solutions for $\bar{R}_{6'}$ on $a=0$ are special solutions for $\bar{R}_6$.

\vspace{12pt}

In section \ref{sec:uni-inners} we will discuss how to reduce the number of parameters appearing in $\Q_1,\Q_2$ by exploiting the inner $\mathfrak{so}(6)$ automorphisms of the algebra and possibly also the residual $\mathfrak{so}(2,4)$ automorphisms. This discussion will  make it easier to identify the allowed solutions to the above quadratic equations that characterise the unimodular extensions.

\subsection{Unimodular extension of the rank-4 and rank-6 cases} \label{sec:uni-ext-rank4-6}
All the rank-4 and rank-6 cases that do not have\footnote{That is, all the extended solutions of section~\ref{sec:summary}, with the extra assumption of setting $a=0$ in $R_{12}$.  Notice, however, that (up to automorphisms) when $a=0$ $R_{12}$ reduces to a special case of $R_7$  and therefore cannot be considered genuinely new.}   $J_{12}$ in $\h$  admit the generalisation of the solution~\eqref{eq:unext-16a-1.1} and~\eqref{eq:unext-16a-1.2} found for rank-2. In particular  we have
\begin{equation}
    \check\xi=0,\qquad 
    s^{\pm,\Is}_{\is,\alpha a}=0,\qquad 
    q^{\pm,\Is}_{\is,1a}=0.
\end{equation}
and the conditions $\{\Q^\Is_\is,\Q^\Js_\js\}=-i\delta^{\Is \Js}\delta_{\is\js}\e$ read
\begin{equation} \label{eq:uni-ext-rank4and6-1}
\begin{gathered}
    \sum_{a}[(q^{+,\Is}_{\is,2a})^2+(q^{-,\Is}_{\is,2a})^2]=\tfrac12,\quad \text{for } \is  =1,2, \ \Is=1,\ldots, N_1,\\
    \sum_a (q^{-,\Is}_{\is,2a}q^{-,\Js}_{\js,2a}+q^{+,\Is}_{\is,2a}q^{+,\Js}_{\js,2a})=0,\quad \text{with } (\Is,\is)\neq (\Js,\js).
    \end{gathered}
\end{equation}
The corresponding $R$-matrices will be denoted by $\bar{R}_7$--$\bar{R}_{11}$ and $\bar{R}_{13}$, $\bar{R}_{14}$.

For a non-trivial extension of the $R_{12}$ solution, denoted by $\bar{R}_{12}$, we must take the generalisation of the solution~\eqref{eq:unext-16a-2.1} and~\eqref{eq:unext-16a-2.2}. We must assume $a \neq 0$ and 
 set
\begin{equation}\label{eq:uniext-r12-1}
    s^{\pm, \Is}_{\is,\alpha a}=0,\qquad 
    q^{\pm, \Is}_{\is,1a}=0,\qquad 
    q^{+,\Is}_{\is,2a}= \epsilon_{\is\js}q^{-,\Is}_{\js,2a}.
\end{equation}
Now the extra conditions
from $\{\Q^\Is_\is,\Q^\Is_\js\}=-i\delta^{\Is\Js}\delta_{\is\js}\e$ are equivalent to
\begin{equation}\label{eq:uniext-r12-2}
\begin{gathered}
    \sum_{\is,a}(q^{-,\Is}_{\is,2a})^2=\tfrac12, \quad \text{for }  \Is=1,\ldots, N_1 ,\\
    \sum_{\is,a} q^{-,1}_{\is,2a}q^{-,2}_{\is,2a}=0,\qquad 
    \sum_{\is,\js,a} \epsilon^{\is\js}q^{-,1}_{\is,2a}q^{-,2}_{\js,2a}=0.
\end{gathered}
\end{equation}

\subsection{Unimodular extensions in the presence of $\mathfrak{so}(6)$ shifts}\label{sec:so6-uni}
As already mentioned, at the level of bosonic $R$-matrices of generic rank, we can always shift $\h$ by an element of $\mathfrak{so}(6)$ and the classical Yang-Baxter equation will still be satisfied. Here we want to explain how to construct unimodular extensions of these shifted $R$-matrices. First of all, by means of $SO(6)$ automorphisms we can always transform a generic $\ts\in \mathfrak{so}(6)$ to a linear combination of the Cartan generators. In our conventions that means
\begin{equation} \label{eq:so6-t-form}
    \ts=\nu_{1}\, \mathcal R_{12}+\nu_{2}\, \mathcal R_{35}+\nu_{3}\, \mathcal R_{46},
\end{equation}
where $\nu_i$ are real coefficients. If $\h\to \h'=\h+\ts$, we must understand how $\ts$ contributes to the equation $[\h',\Q_\is^\Is]=\tfrac12 \Q_\is^\Is -\varepsilon_{\is\js}\check\xi^\Is\,  \Q_\js^\Is$. It is easy to see that $[\ts,\Q_\is^\Is]$ cannot contribute to the term $\tfrac12 \Q_\is^\Is$ on the right-hand-side, so that either it gives something proportional to $\varepsilon_{\is\js}  \Q_\js^\Is$ or it vanishes. In fact, in our basis we can check that $[\ts,\mathcal Q_\pm^{\alpha a}]=i \lambda_a\, \mathcal Q_\mp^{\alpha a}$ (where no summation over $a$ is meant) with the parameters
\begin{equation}
\begin{aligned}
    \lambda_1=-\tfrac12(\nu_1+\nu_2+\nu_3),
    \qquad
    \lambda_2=\tfrac12(\nu_1+\nu_2-\nu_3),\\
    \lambda_3=\tfrac12(-\nu_1+\nu_2+\nu_3), \qquad
    \lambda_4=\tfrac12(\nu_1-\nu_2+\nu_3).
\end{aligned}
\end{equation}
Notice that $[\ts,[\ts,\mathcal Q_\pm^{\alpha a}]]=- (\lambda_a)^2\, \mathcal Q_\pm^{\alpha a}$. The coefficient on the right-hand-side of this relation is negative, which is compatible with the ``square'' of the contribution $\varepsilon_{\is\js}  \Q_\js^\Is$ but not compatible with $\tfrac12 \Q_\is$.

To continue with the discussion, as we learned from the previous computations, we  start from the ansatz $ \Q_\is^\Is=q^{+,\Is}_{\is,\alpha a}\mathcal Q_+^{\alpha a} + i \, q^{-,\Is}_{\is,\alpha a}\mathcal Q_-^{\alpha a}$ for the supercharges. In fact, the contributions from the supercharges $\mathcal S,\overline{\mathcal S}$ are set to zero either by the linear conditions coming from $[\e,\Q_\is^\Is]$ or $[\h,\Q_\is^\Is]$, or by the quadratic relations $\{\Q_\is^\Is,\Q_\js^\Js\}=-i\delta^{\Is\Js}\delta_{\is\js}\e$.
Let us  analyse the case in which $[\ts,\Q_\is^\Is]=-\varepsilon_{\is\js}\Delta \check\xi^\Is\,  \Q_\js^\Is$ with  new parameters $\Delta \check\xi^\Is$. Demanding  $[\ts,[\ts,\Q_\is^\Is]]=-(\Delta\check\xi^\Is)^2\Q_\is^\Is$ then imposes the constraints
\begin{equation}
   [( \lambda_a)^2-(\Delta\check\xi^\Is)^2]\, q^{\pm,\Is}_{\is,\alpha a}=0,\qquad \forall\ \Is,\alpha,a \text{ fixed}.
\end{equation}
An obvious observation is that we must have at least one non-vanishing coefficient among the $q^{\pm,\Is}_{i,\alpha a}$ for each $\Is$, otherwise we would have $\Q_\is^\Is=0$. For the sake of the discussion, let us say that  we have a non-vanishing one for $a=1$. Then we can satisfy the above equations only if $\Delta\check\xi^\Is=\alpha^\Is\lambda_1$, where $\alpha^\Is=\pm 1$. If, for fixed $\Is$, we want more coefficients $q^{\pm,\Is}_{i,\alpha a}$ to be non-zero, for example some with $a=2$, then this will be possible only if $|\lambda_1|=|\lambda_2|$. In other words, this is not possible for the most generic form of $\ts$, but only for special linear combinations of the Cartan generators. A stronger condition will appear when turning on a third index. Notice  that, for fixed $\Is$, it is  possible to have both a non-trivial $\mathfrak{so}(6)$ shift and coefficients with all $a=1,2,3,4$ non-vanishing, because the condition $|\lambda_1|=|\lambda_2|=|\lambda_3|=|\lambda_4|$ can be satisfied by setting 2 $\nu_i$'s to 0, and relating the third one to $\Delta\check\xi^\Is$.

Now, when demanding the  condition $[\ts,\Q_\is^\Is]=-\varepsilon_{\is\js}\Delta \check\xi^\Is\,  \Q_\js^\Is$ we find that it implies
\begin{equation}  \label{eq:cond-off-t}  q^{+,\Is}_{\is,\alpha a}=-\alpha^\Is\epsilon_{\is\js}q^{-,\Is}_{\js,\alpha a}.
\end{equation}
If, before doing the $\mathfrak{so}(6)$ shift, $\check\xi^\Is=0$ then we have two options for each index $\Is$ depending on the sign of $\alpha^\Is$. However, the two signs are related by an automorphism of the superalgebra, and the two choices give rise to the same $R$-matrices. If $\check\xi^\Is\neq 0$ before doing the $\mathfrak{so}(6)$ shift, then we are already imposing a constraint similar to the above one. See for example~\eqref{eq:unext-16a-2.1}, in which case one wants to set $\alpha^\Is=-1$ in order to have compatible constraints. We conclude that it is enough to make sure that the above constraint holds in order to be able to construct unimodular extensions of Jordanian $R$-matrices that contain $\mathfrak{so}(6)$ shifts.

The other option of having $[\ts,\Q_\is^\Is]=0$ can of course be understood from the above discussion when $\Delta\check\xi^\Is=0$. Importantly, in this case the condition~\eqref{eq:cond-off-t} is not necessary anymore, but we must have $\lambda_1=0$, so that we are reducing the possibilities for $\ts$ to be only those that satisfy $\nu_3=-\nu_1-\nu_2$. As before, turning on not only the index $a=1$ but also $a=2$ imposes a stronger constraint on the allowed $\ts$. Notice that turning on 3 indices is not possible because demanding $\lambda_1=\lambda_2=\lambda_3=0$ implies  $\nu_i=0$ for all $i=1,2,3$.

\subsection{Simplifications by inner automorphisms}\label{sec:uni-inners}
The supercharges found in the previous sections still have a large number of free parameters that in fact are not all physical. They can be removed by exploiting inner $\mathfrak{so}(6)$ automorphisms, as well as in principle residual inner automorphisms of $\mathfrak{so}(2,4)$ that leave $\h$ and $\e$ invariant. The latter are summarised for each rank-2 bosonic $R$-matrix admitting a unimodular extension in Table \ref{tab:rank-2-un}.\footnote{Note that in the case of performing an $\mathfrak{so}(6)$ shift, as described in the previous section,  the inner $\mathfrak{so}(6)$ automorphisms have already been exploited in order to transform $\ts$ into the Cartan form \eqref{eq:so6-t-form}. Hence, the corresponding unimodular $R$-matrices can only be simplified with those residual $\mathfrak{so}(6)$ and $\mathfrak{so}(2,4)$ that leave $\h,\e$ and $\ts$ invariant. We will not analyse such simplifications here.}

Before analysing each $R$-matrix $\bar{R}_i$ separately, let us first argue in general how the simplification by $SO(6)$ can be performed.
Consider  a general supercharge $\Q_\is$ of $\mathfrak{psu}(2,2|4)$ of the form \eqref{eq:def-Q-real-coeffs}. We know already that all unimodular extensions require $s^{\pm, \Is}_{\is, \alpha a}=0$ and thus we start from
\begin{equation}
   \Q_\is= q^+_\is \cdot {\cal Q}_+ + i q^-_\is \cdot {\cal Q}_- \ ,
\end{equation}
where for now we will suppress the index $\Is=1, \ldots, N_1$.
There are 15 generators of $\mathfrak{so}(6)$ that we can now exploit to simplify $\Q_\is$. As follows from \eqref{eq:comm-R-QS}, their corresponding adjoint actions are dictated by
\begin{equation}
    [{\cal R}_{AB}, ({\cal Q}_\pm)^{\alpha a}] = \frac{1}{2} (\rho_{AB}^{\text{\tiny AS}})^a{}_b ({\cal Q}^\pm)^{\alpha b} - \frac{1}{2} (\rho_{AB}^{\text{\tiny S}})^a{}_b ({\cal Q}^\mp)^{\alpha b} ,
\end{equation}
with $\rho_{AB}^{\text{\tiny S}}$ ($\rho_{AB}^{\text{\tiny AS}}$) the symmetric (antisymmetric) part of $\rho_{AB}$ as matrices in the $a,b$ indices. In our anti-hermitian matrix realisation of $\mathfrak{so}(6)$ (see appendix \ref{app:psu224}),  there are precisely $\binom{4}{2}=6$ generators for which $\rho_{AB}$ is purely antisymmetric in the $a,b$ indices and that will act as a simple rotation on the $a=1,2,3,4$ of $q^\pm_{\is,\alpha a}$. Let us call  ${\cal R}^{(r)}$ the set of these generators.  The other 9 generators are purely imaginary with  symmetric $\rho_{AB}$ and will thus mix $q^+_{\is,\alpha a}$ and $q^-_{\is,\alpha a}$ with $\is,\alpha$ fixed. They can be divided into 3 generators corresponding to phase transformations rotating $q^+_{\is,\alpha a}$ and $q^-_{\is,\alpha a}$ with also $a$ fixed, whose set we will call ${\cal R}^{(p)}$, and $\binom{4}{2}=6$ generators corresponding to  rotations on the $a$ indices between $q^+_{\is,\alpha a}$ and $q^-_{\is,\alpha a}$, whose set we will call ${\cal R}^{(i)}$. For clarity, we can depict this as
\begin{equation}
    \begin{alignedat}{4}
    {\cal R}^{(r)} &: \qquad && q^\pm_{\is,\alpha a} \leftrightarrow q^\pm_{\is,\alpha b}, \qquad &&\text{with} \qquad && (a,b) \in \{(1,2),(1,3),(1,4),(2,3),(2,4),(3,4)\} , \\
    {\cal R}^{(i)} &: \qquad && q^\pm_{\is,\alpha a} \leftrightarrow q^\mp_{\is,\alpha b}, \qquad &&\text{with} \qquad && (a,b) \in \{(1,2),(1,3),(1,4),(2,3),(2,4),(3,4)\} , \\
    {\cal R}^{(p)} &: \qquad && q^\pm_{\is,\alpha a} \leftrightarrow q^\mp_{\is,\alpha a}, \qquad &&\text{with} \qquad && a \; \text{ fixed} , \\
    \end{alignedat}
\end{equation}
and $\is,\alpha$ fixed in all cases.

Let us now illustrate how these can be used to eliminate some parameters $q^\pm_{\is, \alpha a}$.
We can start focusing, for example, on $q^\pm_{1,1 a}$ and use the $\{(1,2),(1,3),(1,4)\}$ rotations of ${\cal R}^{(r)}$ and ${\cal R}^{(i)}$ to set $q^\pm_{1,1 \{2,3,4\}}=0$ and keep for $(\is, \alpha )=(1,1)$ only $q^\pm_{1,11}$ non-trivial. Notice that the actions of ${\cal R}^{(r)}$ and ${\cal R}^{(i)}$ interfere non-trivially with each other so that on each $(a,b)$ plane the two kinds of rotations must be implemented simultaneously in order to reach the wanted gauge. The other rotations of ${\cal R}^{(r,i)}$ leave this choice invariant and  subsequently the $\{(2,3),(2,4)\}$  can be used to set, for example,  $q^\pm_{1,2 \{3,4\}}=0$ keeping for  $(\is, \alpha)=(1,2)$ only $q^\pm_{1,21}$ and $q^\pm_{1,22}$ non-trivial. Within ${\cal R}^{(r,i)}$ we are finally left with  $(3,4)$, again leaving our previous choices invariant, and which can be used to set, for example, $q^\pm_{2,14}=0$. At last there are also the ${\cal R}^{(p)}$ to exploit. Importantly, so far we have treated the superscripts $+$ and $-$ in the same way (when keeping the remaining indices $\is,\alpha,a$ fixed). This is crucial at the moment of using the ${\cal R}^{(p)}$ transformations, because it means that the previous choices for vanishing coefficients are not spoiled. Considering for example the basis
\begin{equation} \label{eq:generators-so6-phase}
    \mathcal R^{(p)}:\quad \rho^a{}_b \in \left\{ \mathrm{diag}(i,0,0,-i),\mathrm{diag}(0,i,0,-i),\mathrm{diag}(0,0,i,-i) \right\} ,
\end{equation}
we can use the first two elements to set, for example, $q^+_{1,11}=q^+_{1,22}=0$, and the last element (leaving the latter choice invariant) to set, for example, $q^+_{2,13}=0$.

The most convenient ``gauge'' choices to be made will depend on the specific cases under study,  to which we now turn. In particular,  the type of constraints following from the  (anti)-commutation relations required for the unimodular  extensions  will play an important role in this. Note  that these constraints will of course be left invariant after the action of inner automorphisms that leave $\h$ and $\e$ invariant.
If possible, we may also employ the residual inner $\mathfrak{so}(2,4)$ automorphisms of each case. We first consider the rank-2 extensions and comment on the possible higher-ranks at the end. 

\subsubsection*{$\bar{R}_1$ and $\bar{R}_{1'}$}
Given the results of section \ref{sec:rank2-uniext}, we  could  distinguish here two cases, i.e.~$a$ generic or $a=0$. In both cases, only  the   parameters $q^\pm_{\is, 2a}$ were left non-vanishing after imposing the required (anti)commutation relations. With the exception of $J_{12}$ (which acts as a phase transformation $q^\pm_{\is,\alpha a} \leftrightarrow q^\mp_{\is,\alpha a}$ with fixed indices $\is,\alpha$ and $a$), the inner automorphisms of Table \ref{tab:rank-2-un} then all act trivially on the supercharges and therefore their possible usage can be discarded. 

$\bullet$  $\bar{R}_{1}$.  In this case, the parameters $q^\pm_{\is, 2a}$ are subjected to the  constraints \eqref{eq:unext-16a-1.2}. As illustrated above, we can use $\{(1,2),(1,3),(1,4)\}$ of ${\cal R}^{(r,i)}$ to keep only $q^\pm_{1,21}$ and $q^\pm_{2,2a}$. Next, we can use $\{(2,3),(2,4)\}$ of ${\cal R}^{(r,i)}$ to keep only $q^\pm_{1,21}$, $q^\pm_{2,21}$, and $q^\pm_{2,22}$. At last, we can use the first two elements of \eqref{eq:generators-so6-phase} to reduce this to $q^+_{1,21}$, $q^\pm_{2,21}$, and $q^+_{2,22}$. The constraints \eqref{eq:unext-16a-1.2} are then solved by $q^+_{2,21}=0$ and\footnote{The quadratic equation admits also the solution for $q^+_{1,21}$ with the negative sign. A change of sign on $\Q_1$ or $\Q_2$ however leaves the $R$-matrix \eqref{eq:rmatrix-rank2-uniext} invariant.}
\begin{equation}\label{eq:end-solution-rank2-uni-r1.2}
    q^+_{1,21} = \frac{1}{\sqrt{2}}, \qquad  (q^+_{2,22})^2+(q^-_{2,21})^2=\frac{1}{2} \ .
\end{equation}
 Note that, the remaining $(3,4)$ of ${\cal R}^{(r,i)}$ and the last element of \eqref{eq:generators-so6-phase} do not act on these parameters. These transformations can be used to simplify higher-rank extensions, as we will do below starting from $R_7$. Similarly, one can check that the residual $J_{12}$ action on our reduced parameter space does not  simplify this case further, in fact the action of $J_{12}$ is not independent from  that of the first two elements of \eqref{eq:generators-so6-phase}. 

$\bullet$ $\bar{R}_{1'}$. In this case,   the   parameters $q^\pm_{\is, 2a}$ are subjected to different constraints, namely $q^+_{\is,2a}=\epsilon_{\is\js} q^-_{\js, 2a}$ and \eqref{eq:unext-16a-2.2}. As before, we can now exploit the $\{(1,2),(1,3),(1,4)\}$ of ${\cal R}^{(r,i)}$   to set $q^\pm_{1,2\{2,3,4\}}=0$. The constraint $q^+_{\is,2a}=\epsilon_{\is\js} q^-_{\js, 2a}$ then actually implies that only the 4 coefficients $q^\pm_{\is,21}$ are left non-trivial and they are related as $q^+_{1,21}=q^-_{2,21}, q^+_{2,21}=-q^-_{1,21}$. After this we still have the freedom $\{(2,3),(2,4),(3,4)\}$  of ${\cal R}^{(r,i)}$. In this case, they are not useful because they rotate among each other coefficients that are already vanishing. One could imagine using them to eliminate  6 more parameters of a  higher-rank unimodular extension, however this case (i.e.~the parameter $a$ generic)  does not admit such extensions.   The first element of \eqref{eq:generators-so6-phase} can subsequently be used to set, for example, $q^+_{2,21}=0$  which upon the constraints also implies $q^-_{1,21}=0$. We are left with one parameter which must be fixed by solving  \eqref{eq:unext-16a-2.2}. We take\footnote{Again the choice of  sign on $q^-_{2,21}$ would not affect the resulting $R$-matrix.}
    \begin{equation} \label{eq:end-solution-rank2-uni-r1.1}
        q^-_{2,21}=q^+_{1,21}=\frac{1}{\sqrt{2}} \ .
    \end{equation}
 Note that here we did not need  the residual  $J_{12}$ automorphisms. In fact, it does not leave the choice $q^+_{1,21}=q^-_{2,21}=0$ invariant. 

Let us remark that in the special case for $\bar{R}_1$ with $q^+_{2,22}=0$, the solution \eqref{eq:end-solution-rank2-uni-r1.2} reduces to \eqref{eq:end-solution-rank2-uni-r1.1} and thus the latter holds for truly generic $a$, as it should.


\subsubsection*{$\bar{R}_2$, $\bar{R}_{2'}$, $\bar{R}_3$, $\bar{R}_4$,  $\bar{R}_5$, and $\bar{R}_{5'}$}
All these cases are  analogous to the case $\bar{R}_1$ and $\bar{R}_{1'}$. The $\bar{R}_2$, $\bar{R}_3$, $\bar{R}_4$, and $\bar{R}_5$ are solved by \eqref{eq:end-solution-rank2-uni-r1.2}, while $\bar{R}_{2'}$ and $\bar{R}_{5'}$ are solved by \eqref{eq:end-solution-rank2-uni-r1.1},   with all the other parameters  set to zero. Note, however, that certain versions of the $\bar{R}_{5'}$ have a corresponding higher-rank extension, i.e.~$\bar{R}_{12}$, for which we will be able to use the remaining $SO(6)$ transformations that have not been exploited in the $\bar{R}_{1'}$ case (see below).

\subsubsection*{$\bar{R}_6$ and $\bar{R}_{6'}$}

Again we must distinguish $a=0$ or $a$ generic. In both cases, all the $q^\pm_{\is,\alpha a}$ were left non-vanishing.

$\bullet$ $\bar{R}_{6}$.  In this case, the parameters $q^\pm_{\is,\alpha a}$ are subjected to \eqref{eq:quadr6a0}. Looking at table \ref{tab:rank-2-un} we have, besides the $SO(6)$ automorphisms to exploit, also residual $SO(2,4)$ automorphisms generated by $J_{12}$, $J_{13}$ and $J_{23}$. Here, $J_{13}$ and $J_{23}$ are like $(r,i)$ rotations, but solely on the $\alpha$ indices, i.e.~they act as $q^\pm_{\is,\alpha a} \leftrightarrow q^\pm_{\is,\beta a}$ and $q^\pm_{\is,\alpha a} \leftrightarrow q^\mp_{\is,\beta a}$ for fixed $\is, a$ and for $(\alpha, \beta) = (1,2)$. The $J_{12}$ automorphism acts like the phase rotations, though it will act as $q^\pm_{\is, \alpha a} \leftrightarrow q^\mp_{\is, \alpha a}$ for all $\is,\alpha, a$ simultaneously. With the same machinery as above, it is not hard to show that all of these automorphisms combined allow one to set the following parameters (18 in total) of the supercharges to zero: 
\begin{equation}
    q^\pm_{1,1\{2,3,4\}} = q^\pm_{1,2\{1,3,4\}}= q^\pm_{2,14} = q^+_{1,11} = q^+_{1,22} = q^+_{2,21}=  q^-_{2,23} =0 \ .
\end{equation}
The quadratic constraints \eqref{eq:quadr6a0} will then further imply that\footnote{Here we again made a choice in sign, which  is compatible with the special solution \eqref{eq:r6specialsol} below, and which will be related by an inner automorphism. See comments around \eqref{eq:r6specialsol}.}
\begin{equation}
    q^-_{1,11}=q^-_{1,22}=\frac{1}{2}, \qquad q^-_{2,11}=q^-_{2,22}=q^+_{2,12}=0, \qquad q^-_{2,21}=- q^-_{2,12}, 
\end{equation}
and
\begin{equation} \label{eq:quadr6a0simp}
    \begin{aligned}
        &(q^+_{2,22})^2 +(q^+_{2,23})^2 + (q^+_{2,24})^2 + (q^-_{2,21})^2 + (q^-_{2,24})^2 = \frac{1}{4} , \\
        &(q^+_{2,11})^2 + (q^+_{2,13})^2 + (q^-_{2,12})^2 + (q^-_{2,13})^2 = \frac{1}{4} , \\
        & (q^+_{2,11}+q^+_{2,22}) q^-_{2,21} - q^-_{2,13} q^+_{2,23} =0 , \\
        &  q^+_{2,13} q^+_{2,23} =0 . 
    \end{aligned}
\end{equation}
This branches out in several possibilities.
After exploiting all of the (residual) automorphisms, we thus effectively have 8 remaining  parameters subjected to 4 quadratic constraints.

$\bullet$ $\bar{R}_{6'}$. In this case, the $q^\pm_{\is,\alpha a}$ are subjected to $q^+_{\is, \alpha a} = (-1)^\alpha \epsilon_{\is\js} q^-_{\js,\alpha a}$ and \eqref{eq:quadp0p3}. Following now precisely the illustration of the beginning of this section, we can use all of the  $\mathfrak{so}(6)$ elements to set $q^\pm_{1,1\{2,3,4\}}=q^\pm_{1,2\{3,4\}}=q^\pm_{2,14}=q^+_{1,11}=q^+_{1,22}=q^+_{2,13}=0$. The constraint $q^+_{\is, \alpha a} = (-1)^\alpha \epsilon_{\is\js} q^-_{\js,\alpha a}$ now further implies that the only non-vanishing parameters are $q^\pm_{1,21}=\pm q^\mp_{2,21}$,  $q^-_{1,11}=q^+_{2,11}$, and $q^-_{1,22}= - q^+_{2,22}$. This, in fact, suffices to completely solve \eqref{eq:quadp0p3}. We find that we must set $q^\pm_{1,21}=0= q^\mp_{2,21}$ and\footnote{When solving the quadratic equations we actually find $q^-_{1,11}=q^+_{2,11}=\pm \frac{1}{2},  q^-_{1,22}=- q^+_{2,22}= \pm \frac{1}{2}$ where the two choices of signs are uncorrelated. However, we first notice that the 4 choices of combinations of signs lead to only 2 independent solutions for the $R$-matrix, namely when the signs either agree or are opposite. This is due to the fact that changing the overall sign of $\Q_1$ or $\Q_2$ does not change the $R$-matrix. Nevertheless, we find that these two seemingly distinct choices (consider $\Q_\is$ with $q^-_{1,11}=q^-_{1,22}=1/2$ and   $\Q'_\is$ with $q^-_{1,11}=-q^-_{1,22}=1/2$) are related by an inner automorphism $\Q_\is=M \Q'_\is M^{-1}$ with $M\in PSU(2,2|4)$. In fact several  elements $M\in PSU(2,2|4)$ realise this, including first swapping  $\Q_1\leftrightarrow \Q_2$ and then acting with $J_{12}$ (which effectively exchanges the superscripts $+$ and $-$).}
\begin{equation} \label{eq:r6specialsol}
    q^-_{1,11}=q^+_{2,11}= q^-_{1,22}=- q^+_{2,22}=\frac{1}{2}  \ .
\end{equation}
All the parameters are determined, and we did not need to use the residual $\mathfrak{so}(2,4)$ automorphisms. Interestingly,  note that even though we exploited the $(3,4)$ of ${\cal R}^{(r,i)}$ and the latter element of \eqref{eq:generators-so6-phase}, they  are restored as inner automorphisms in this case because of the constraint $q^+_{\is, \alpha a} = (-1)^\alpha \epsilon_{\is\js} q^-_{\js,\alpha a}$.

\subsubsection*{$\bar{R}_7$, $\bar{R}_8$, $\bar{R}_9$, $\bar{R}_{10}$, and $\bar{R}_{11}$}
These $R$-matrices originate from  bosonic rank-4 $R$-matrices and they thus have the $Q_\is^\Is$ with $\is=1,2$ and $\Is=1,2$. Their only non-vanishing parameters after the required commutation relations are $q^{\pm, \Is}_{\is, 2a}$ subjected to the constraints \eqref{eq:uni-ext-rank4and6-1}. As we know from section \ref{sec:uni-ext-rank4-6}, these cases are  similar to the unimodular extension  $\bar{R}_1$, and in fact using $\{(1,2),(1,3),(1,4),(2,3),(2,4)\}$  of ${\cal R}^{(r,i)}$ and  the first two elements of \eqref{eq:generators-so6-phase} as before, the equations \eqref{eq:uni-ext-rank4and6-1} for $q^{\pm, 1}_{\is, 2a}$ can be solved as in \eqref{eq:end-solution-rank2-uni-r1.2}, i.e.~$q^{+,1}_{2,21}=0$ and
\begin{equation}
    q^{+,1}_{1,21} =  \frac{1}{\sqrt{2}},  \qquad (q^{+,1}_{2,22})^2+(q^{-,1}_{2,21})^2=\frac{1}{2} \ .
\end{equation}
Now we can focus on $q^{\pm, 2}_{\is, 2a}$ and use the remaining $\mathfrak{so}(6)$ elements to simplify them. First, with $\{(3,4)\}$  of ${\cal R}^{(r,i)}$ we can set $q^{\pm,2}_{1,24}=0$. The last  element of \eqref{eq:generators-so6-phase} can be used to set $q^{-,2}_{1,23}=0$.  Note that we do not have further inner automorphisms of $\mathfrak{so}(2,4)$ that leave $\{\h,\e,\e_+,\e_-\}$ invariant and that would act non-trivially on the supercharges. One can now check that \eqref{eq:uni-ext-rank4and6-1} further enforces $q^{+,2}_{\is,21}=0$ and leaves us with 13 parameters subjected to 6 constraints. We checked that these can admit real solutions with 7 remaining free parameters in the supercharges $\Q_\is^\Is$. We will not analyse them further here.

\subsubsection*{$\bar{R}_{12}$}
Recall that for a genuinely new unimodular extension $\bar{R}_{12}$ we must assume $a\neq 0$ and thus this will be similar to the $\bar{R}_{1'}$ case. After the required (anti)-commutation relations, the non-vanishing parameters are $q^{\pm, \Is}_{\is, 2a}$ subjected to $q^{+, \Is}_{\is, 2a}=\epsilon_{\is\js} q^{-, \Is}_{\js, 2a}$ and \eqref{eq:uniext-r12-2}. We exploit the $\{(1,2),(1,3),(1,4)\}$ of ${\cal R}^{(r,i)}$ as well as the first element of \eqref{eq:generators-so6-phase}  to set $q^{\pm,1}_{1,2\{2,3,4\}}=0$ and $q^{-,1}_{2,21}=0$. The constraints for $\Is=1$ then completely fix $Q^1_\is$ and in particular imply that also $q^{\pm,1}_{2,2\{2,3,4\}}=0$ and $q^{+,1}_{1,21}=0$ as well as
\begin{equation}
    q^{-,1}_{1,21}=-q^{+,1}_{2,21}=\frac{1}{\sqrt{2}} \ ,
\end{equation}
where we made an inconsequential  choice of the sign that will not affect the $R$-matrix. Now we consider $\Is=2$ and possibly exploit the remaining $SO(6)$ transformations, i.e.~$\{(2,3),(2,4),(3,4)\}$ of ${\cal R}^{(r,i)}$ as well as the latter two elements of \eqref{eq:generators-so6-phase}.  This allows to set respectively $q^{\pm,2}_{1,2\{3,4\}}=q^{\pm,2}_{2,24}=0$ and $q^{-,2}_{2,22}=0$ which implies using $q^{+, \Is}_{\is, 2a}=\epsilon_{\is\js} q^{-, \Is}_{\js, 2a}$ that also  $q^{\pm,2}_{2,23}=0$ and $q^{+,2}_{1,22}=0$.\footnote{In fact, we did not need to use the latter element of \eqref{eq:generators-so6-phase} which remains as a freedom that does not act on the surviving parameters.} Hence at this stage only  $q^{-,2}_{1,21}=-q^{+,2}_{2,21}$, $q^{-,2}_{2,21}=q^{+,2}_{1,21}$, and $q^{-,2}_{1,22}=-q^{+,2}_{2,22}$ are undetermined.
Using the results of $Q^1_\is$, the remaining constraints simply imply $q^{-,2}_{1,21}=q^{-,2}_{2,21}=q^{+,2}_{2,21}=q^{+,2}_{1,21}=0$ and
\begin{equation}
    q^{-,2}_{1,22} =-q^{+,2}_{2,22}=  \frac{1}{\sqrt{2}} \ ,
\end{equation}
where again we made a choice for the sign.
Concluding, for the unimodular extension  $\bar{R}_{12}$  we do not have any remaining free parameters in the supercharges.

\subsubsection*{$\bar{R}_{13}$ and $\bar{R}_{14}$}

These  $R$-matrices originate from the bosonic  rank-6 $R$-matrices and are similar to the unimodular extension $\bar{R}_1$, albeit now with supercharges $Q^\Is_\is$ with  $\is=1,2$ and $\Is=1,2,3$. While the $\bar{R}_1$ (from rank-2) had one remaining free parameter, their rank-4 generalisations admit solutions with 7 remaining free parameters (see $\bar{R}_7$--$\bar{R}_{11}$ discussed above) after using residual inner automorphisms. The rank-6 generalisation $\bar{R}_{13}$ and $\bar{R}_{14}$ (where no further inner automorphisms can be exploited)  will have in principle 16 more initial parameters than the rank-4  cases which all have to satisfy the same 10 constraints as before, i.e.~\eqref{eq:uni-ext-rank4and6-1}. 
Because of  these reasons, it is clear that the $\bar{R}_{13}$ and $\bar{R}_{14}$ have a complicated and large system to solve with a large number of remaining free parameters.   We will not analyse this further.


\section{Preserved isometries and superisometries} \label{sec:siso}

In this section, we consider the unimodular extensions of the rank-2 $R$-matrices with $\h,\e\in\mathfrak{so}(2,4)$ and identify the generators of $T_{\bar{A}}\in \mathfrak{psu}(2,2|4)$ whose adjoint  action commutes with the action of $\bar{R}$, i.e.
\begin{equation} \label{eq:cond-siso}
    \mathrm{ad}_{T_{\bar{A}}} \ \bar{R} = \bar{R} \ \mathrm{ad}_{T_{\bar{A}}} \ .
\end{equation}
Because of the Jacobi identity, these generators span a subalgebra of $\mathfrak{psu}(2,2|4)$.
They have the important interpretation that they correspond to  (super)isometries of the supergravity background. Indeed, at the group level, they generate global left transformations $g\rightarrow g_L g$ with $g_L \in PSU(2,2|4)$ constant and satisfying
\begin{equation}
    \mathrm{Ad}^{-1}_{g_L} \ \bar{R} \ \mathrm{Ad}_{g_L}  = \bar{R} \ ,
\end{equation}
which leaves the deformed semisymmetric coset sigma-model action \eqref{eq:hyb-ssssm} invariant.

\begin{table}[h!]
\centering
\begin{adjustbox}{center}
    \begin{tabular}{l|c|l|c}
    $\bar{R}$ & \multicolumn{1}{|c|}{conditions} & \multicolumn{1}{|c|}{$T_{\bar{A}}\in \mathfrak{so}(2,4)\oplus \mathfrak{so}(6)$}   & supercharges  \\
    \hline

    \hline
    \multirow{8}{*}{1} 
     & ($a=0$ for all $\bar R_1$) & $D+J_{03}$, $p_+$, $J_{12}-\mathcal R_{46}$,   $\mathfrak{k}_2 (w)$ &  0\\
    \cline{2-4}
     & $b=-1$ & $D+J_{03}$, $p_+$, $p_1$, $p_2$, $J_{12}-R_{46}$,  $\mathfrak{k}_2 (w)$  &  0\\
    \cline{2-4}
     &   $b=0$ & $D+J_{03}$, $p_+$, $j_{013}$, $j_{023}$, $J_{12}-\mathcal R_{46}$, $\mathfrak{k}_2 (w)$ &  0\\
    \cline{2-4}
     &   $b=-1/2$   & $D+J_{03}$, $p_0$, $p_3$, $k_0+k_3$, $J_{12}-\mathcal R_{46}$, $\mathfrak{k}_2 (w)$  &    8\\
    \cline{2-4}
     &  $q^+_{2,22}=0$ & $D+J_{03}$, $p_+$, $J_{12}$,   $\mathfrak{k}_1$  &  0\\
    \cline{2-4}
     &   $b=-1$, $q^+_{2,22}=0$ & $D+J_{03}$, $p_+$, $p_1$, $p_2$, $J_{12}$,  $\mathfrak{k}_1$  &  0\\
    \cline{2-4}
     &   $b=0$, $q^+_{2,22}=0$ & $D+J_{03}$, $p_+$, $j_{013}$, $j_{023}$, $J_{12}$,  $\mathfrak{k}_1$  &  0\\
    \cline{2-4}
     &   $b=-1/2$, $q^+_{2,22}=0$   & $D+J_{03}$, $p_0$, $p_3$, $k_0+k_3$, $J_{12}$, $\mathfrak{k}_1$  &    12\\
    \hline

    \hline
    \multirow{2}{*}{1'} & ($a\neq 0$ for all $\bar R_{1'}$) &  $D+J_{03}$, $p_+$, $J_{12}$, $\mathfrak{k}_1$ & 0 \\
    \cline{2-4}
     &  $b=-1/2$ &  $D+J_{03}$, $p_0$, $p_3$, $k_0+k_3$, $J_{12}$, $\mathfrak{k}_1$ & 0 \\
    \hline
    
    \hline
    \multirow{2}{*}{2} &  ($a=0$ for all $\bar R_{2}$) &  $p_0$, $p_3$, $J_{12}-\mathcal R_{46}$,   $\mathfrak{k}_2 (w)$ & 4 \\
    \cline{2-4}
     &  $q^+_{2,22}=0$ &  $p_0$, $p_3$, $J_{12}$,  $\mathfrak{k}_1$ & 6 \\
    \hline

    \hline
    \multirow{1}{*}{2'} & $a\neq 0$ &  $p_0$, $p_3$, $J_{12}$, $\mathfrak{k}_1$  & 0 \\
    \hline

    \hline
    \multirow{2}{*}{3} & $-$ & $p_+$, $p_1$, $p_2$, $\mathfrak{k}_2 (w)$ &  0\\
    \cline{2-4}
     & $q^+_{2,22}=0$ & $p_+$, $p_1$, $p_2$,$\mathfrak{k}_1$ & 0  \\
    \hline

    \hline
    \multirow{2}{*}{4} & $-$ & $p_+$, $j_{013}$, $j_{023}$, $\mathfrak{k}_2 (w)$ & 0 \\
    \cline{2-4}
     & $q^+_{2,22}=0$ & $p_+$, $j_{013}$, $j_{023}$, $\mathfrak{k}_1$ & 0 \\
    \hline

    \hline
    \multirow{2}{*}{5} & ($a=0$ for all $\bar R_5$) & $p_+$, $k_0+k_3+2p_3$, $J_{12}-\mathcal R_{46}$, $\mathfrak{k}_2 (w)$ & 0 \\
    \cline{2-4}
     &  $q^+_{2,22}=0$ & $p_+$, $k_0+k_3+2p_3$, $J_{12}$, $\mathfrak{k}_1$  & 0 \\
    \hline

    \hline
    \multirow{1}{*}{5'} & $a\neq 0$ & $p_+$, $k_0+k_3+2p_3$, $J_{12}$, $\mathfrak{k}_1$ & 0 \\
    \hline

    \hline
    \multirow{4}{*}{6} &
     ($a=0$ for all $\bar R_6$)& $p_0$, $\mathfrak u(1)$, $\mathfrak u(1)$  & 0\\
    \cline{2-4} & $q^\pm_{\is,\alpha \{3,4\}}=0, q^-_{2,21}\neq 0$& $p_0$, $J_{13} -w^{-1} J_{12}$, $\mathfrak{k}_2 (-w^{-1})$  & 0\\
    \cline{2-4} & $ q^\pm_{\is,\alpha \{3,4\}}=0, q^+_{2,11}=-q^+_{2,22}=\pm \tfrac12$& $p_0$, $J_{12}$, $\mathfrak k_3$  & 0\\
    \cline{2-4} & $ q^\pm_{\is,\alpha \{3,4\}}=0, q^+_{2,11}=q^+_{2,22}=\pm \tfrac12$& $p_0$, $J_{12}+\mathcal R_{35},J_{13}-\mathcal R_{25},J_{23}+\mathcal R_{23}$, $\mathfrak k_4$  & 0\\
    \hline

    \hline
    6' & $a\neq 0$& $p_0$, $J_{12}$, $\mathfrak{k}_3$  & 0\\
      \hline
    \end{tabular}
\end{adjustbox}
    \caption{The bosonic generators $T_{\bar{A}}\in \mathfrak{so}(2,4)\oplus \mathfrak{so}(6)$, as well as the number of fermionic elements in $\mathfrak{psu}(2,2|4)$ that satisfy \eqref{eq:cond-siso} for  the unimodular extended rank-2 Jordanian $R$-matrices of the form ${r}=\h\wedge \e -\frac{i}{2} (\Q_1\wedge \Q_1 + \Q_2 \wedge \Q_2)$. Such generators represent the residual (super)isometries of the deformed supergravity background. If a parameter is not specified, it is assumed to be generic (modulo constraints such as \eqref{eq:end-solution-rank2-uni-r1.2}). The algebras $\mathfrak{k}_1$, $\mathfrak{k}_2 (w)$, $\mathfrak{k}_3$, and $\mathfrak{k}_4$ are subalgebras of $\mathfrak{so}(6)$ and defined in \eqref{eq:k1}, \eqref{eq:k2}, \eqref{eq:k3}, and \eqref{eq:k4} respectively. Here, $w$ is the continuous parameter $q^-_{2,21}/q^+_{2,22}$.
    To save space we are using the shorthand notations $p_\pm=p_0\pm p_3$ and $j_{\mu\nu\rho}=J_{\mu\nu}-J_{\nu\rho}$.
    }
    \label{tab:rank-2-siso-full}
\end{table}

We collect our results in Table \ref{tab:rank-2-siso-full}, which is the complete version of Table \ref{tab:rank-2-siso} of our summary section. Because the $R$-matrix preserves the degree of the superalgebra element, we can consider bosonic and fermionic generators $T_{\bar{A}} \in \mathfrak{psu}(2,2|4)$ separately. For the fermionic generators, we only list the number of independent generators, corresponding to the number of superisometries preserved. We also use a short-hand notation for those elements $T_{\bar{A}}$ that are strictly in $\mathfrak{so}(6)$, as they are repeated frequently. We define a 9-dimensional algebra, 
\begin{equation} \label{eq:k1}
    \mathfrak{k}_1 \equiv \mathrm{span}(\mathcal R_{16}-\mathcal R_{24}, \mathcal R_{14}+\mathcal R_{26},\mathcal R_{36}+\mathcal R_{45}, \mathcal R_{34}+\mathcal R_{56}, \mathcal R_{13}+\mathcal R_{25},\mathcal R_{15}-\mathcal R_{23} ,\mathcal R_{12}, \mathcal R_{35}, \mathcal R_{46}) \ ,
\end{equation}
which is the algebra appearing once the supercharges $\Q_\is$ are completely determined, with the exception of the distinct $\bar{R}_{6}$ and $\bar{R}_{6'}$ cases. 
The elements of $\mathfrak{k}_1$ correspond  to  the remaining $\mathfrak{so}(6)$ automorphisms that were not exploited in the previous section, i.e.~the $\{(2,3),(2,4),(3,4)\}$ of ${\cal R}^{(r,i)}$ as well as the last  two   phase transformations of \eqref{eq:generators-so6-phase}, in addition to the first phase transformation of \eqref{eq:generators-so6-phase}.\footnote{Note that the requirement of an inner automorphism that leaves $\{\h,\e, Q_\is\}$ invariant, i.e.~$[x,\h]=[x,\e]=[x,Q_\is]=0$, is stronger than the requirement of a (super)isometry.} It is not hard to show that the algebra $\mathfrak{k}_1$ is isomorphic to $\mathfrak{su}(3)\oplus \mathfrak{u}(1)$ by identifying a central element and calculating  the dual Coxeter number of the remaining 8-dimensional algebra. We also define the following two 4-dimensional algebras
\begin{align}
    \mathfrak{k}_2 (w) &\equiv \mathrm{span}(\mathcal R_{13}+\mathcal R_{25}, \mathcal R_{15}-\mathcal R_{23}, \mathcal R_{12}-\mathcal R_{35}, \mathcal R_{13}+ w \mathcal R_{35}) \simeq  \mathfrak{su}(2)\oplus \mathfrak{u}(1), \label{eq:k2}\\
    \mathfrak{k}_3 &\equiv \mathrm{span}(\mathcal R_{13}+\mathcal R_{25}, \mathcal R_{15}-\mathcal R_{23}, \mathcal R_{12}-\mathcal R_{35}, \mathcal R_{12}+ \mathcal R_{35}) \simeq \mathfrak{su}(2)\oplus \mathfrak{u}(1),  \label{eq:k3}
\end{align}
where $w = q^-_{2,21}/q^+_{2,22}$ with the assumption $q^+_{2,22}\neq 0$. The algebra $\mathfrak{k}_2 (w)$ appears for those cases in which the supercharges $\Q_\is$ have a free parameter, while $\mathfrak{k}_3$ is the subalgebra of isometries of $\mathfrak{so}(6)$ for $\bar{R}_{6'}$. The first three elements of $\mathfrak{k}_2 (w)$ and $\mathfrak{k}_3$ correspond to the $(3,4)$ of ${\cal R}^{(r,i)}$ and the latter element of \eqref{eq:generators-so6-phase}, which are all residual $\mathfrak{so}(6)$ inners of the respective cases.  Furthermore, we remark that even though the algebra $\mathfrak{k}_2 (w)$ appears to depend on a continuous parameter, its structure constants are $w$-independent once we redefine $\mathcal R_{13}+ w \mathcal R_{35}\to \mathcal R_{13}+ w \mathcal R_{35} +\frac{w}{2}(\mathcal R_{12}-\mathcal R_{35})$. A further appropriate shift of this element with $\mathcal R_{13}+\mathcal R_{25}$ exposes that $\mathfrak{k}_2 (w) \simeq \mathfrak{su}(2)\oplus \mathfrak{u}(1)$. How this algebra is embedded in $\mathfrak{so}(6)$, however, does depend on $w$. 

From the point of view of the sigma-model action, the extra  elements in $\mathfrak{k}_1, \mathfrak{k}_2, \mathfrak{k}_3$ on top of the  residual $\mathfrak{so}(6)$  inners seem to imply the possibility of further simplifications of the supercharges. Nevertheless,   these transformations do not act independently on the reduced parameter space  or compared to the $J_{12}$ action.

The (super)isometries of the $\bar R_6$ matrix deserve a separate comment because of the presence of a large set of free parameters in the solution. For generic values of the free coefficients $q^\pm_{2,\alpha a}$ there are no superisometries and the bosonic isometries are given just by $p_0$ and two more commuting generators: one of them is a combination of Lorentz generators with elements from $\mathfrak{so}(6)$, while the other belongs to $\mathfrak{so}(6)$ only. We do not write explicitly such linear combinations of generators because they depend non-trivially on the free coefficients $q^\pm_{2,\alpha a}$. Nevertheless, we obviously have symmetry enhancements for special values of the coefficients $q^\pm_{2,\alpha a}$. In the table we do not list all possible cases but just the ones that present the largest possible isometries. They all correspond to setting $q^\pm_{2,\alpha a}=0$ for $a=3,4$, plus other possible constraints in order to solve the quadratic conditions~\eqref{eq:quadr6a0simp}. In particular, we see that  assuming $q^-_{2,21}\neq 0$ implies that we have bosonic isometries generated by $J_{13} -w^{-1} J_{12}$ as well as $\mathfrak k_2(-w^{-1})$.
Other bosonic isometries are found when assuming $q^-_{2,21}=0$. In particular, we also defined the $\mathfrak{so}(6)$ subalgebra
\begin{equation}
\label{eq:k4}
    \mathfrak k_4\equiv \text{span}(\mathcal R_{13}+\mathcal R_{25},\mathcal R_{15}-\mathcal R_{23},\mathcal R_{12}-\mathcal R_{35},\mathcal R_{46})\simeq \mathfrak{su}(2)\oplus \mathfrak{u}(1).
\end{equation}
None of these special cases have residual superisometries, allowing us to conclude that this will be the case for any value of the free parameters.

Due to the large number of remaining free parameters, we will not do a full analysis of the preserved (super)isometries of the  unimodular extensions of the higher-rank solutions. An exception is $\bar{R}_{12}$ for which all the parameters in $\Q_{\is}$ are completely determined, and we find for any value of $a\neq 0$ that there are no residual superisometries. For the other unimodular higher-rank $R$-matrices it is natural to expect that the superisometries will also be completely broken for a generic choice of the parameters.

Before concluding this section, let us point out that the $\mathfrak{so}(6)$ isometries in Table~\ref{tab:rank-2-siso-full} confirm the compatibility between our findings for the unimodular extensions of $R$-matrices with $\mathfrak{so}(6)$ shifts (see section~\ref{sec:so6-uni}) and the interpretation of these deformations as compositions of a Jordanian deformation on $AdS$ followed by a TsT deformation along  $\ts$ and $\e$. Notice that this construction  is always possible because $\e$ is  an isometry after implementing the Jordanian deformation. Looking at Table~\ref{tab:rank-2-siso-full} we see that the $\mathfrak{so}(6)$ subalgebra of isometries preserved by the unimodular $R$-matrices is either of rank 3 (that is the case of $\mathfrak k_1$ which has 3 Cartans) or rank 2 (that is the case of $\mathfrak k_2, \mathfrak k_3, \mathfrak k_4$ which have 2 Cartans) or rank 1 (the generic case of $\bar R_6$). Reading the Table we see that every time we set $q^+_{2,22}=0$ (implying that $q^\pm_{\is,\alpha a}\neq 0$ only for $a=1$) there is an enhancement of the symmetry to $\mathfrak k_1$. That means that we can implement a TsT with the most generic $\ts$, and this is compatible with what we found in section~\ref{sec:so6-uni}: if we only have coefficients with $a=1$, then $\ts$ is unrestricted.
Moreover, in the table we also see that when we have both $a=1$ and $a=2$ turned on then the symmetry algebra is $\mathfrak k_2$ or $\mathfrak k_3$, which means that $\ts$ can only be a combination of just 2 Cartans. This is again compatible with what we found in section~\ref{sec:so6-uni}.


\section{Conclusions} \label{sec:conclusions}

We have classified all antisymmetric bosonic Jordanian solutions of the classical Yang-Baxter equation on $\mathfrak{psu}(2,2|4)$ and constructed their most generic fermionic extensions that ensure unimodularity. These properties are significant for constructing a Yang-Baxter deformation of the integrable string sigma-model on $\mathfrak{psu}(2,2|4)$ which gives rise to the maximally supersymmetric $AdS_5\times S^5$ background. In particular, antisymmetric solutions of the CYBE ensure that the deformation preserves the property of integrability. Unimodularity, in addition, ensures that the deformed $AdS_5\times S^5$ background will still solve the type IIB supergravity equations of motion. 

When the bosonic Jordanian $R$-matrices are not extended with fermionic supercharges, they are non-unimodular and the corresponding background  solves the modified (or generalised) supergravity equations. We find that they are at most of rank-6, where the rank denotes the number of bosonic elements in the construction of the $R$-matrix. For all these cases, we analyse whether or not they admit a unimodular extension, which is a stringent requirement for rank-2, but is always possible for rank-4 and rank-6. For the simplest unimodular extensions, namely those of rank-2,  we also analyse the preserved (super)isometries of the corresponding deformed supergravity background and find that they preserve at most 12 superisometries. All of our main results are structurally summarised in section \ref{sec:summary}.

A peculiarity of all Jordanian deformations (both unimodular and non-unimodular) is that the overall deformation parameter sitting in front of the $R$-matrix can be fixed to a finite (non-vanishing) value by means of a field redefinition. This can be seen explicitly by implementing an inner automorphism generated by $\h$ on the generic extended $R$-matrix in~\eqref{eq:ext-r-Tol}, which would rescale it by an overall coefficient. The possibility of fixing the deformation parameter is related to the reinterpretation of homogeneous Yang-Baxter deformations as deformations of non-abelian T-duality, and to the fact that the 2-cocycles of Jordanian deformations are in fact coboundaries. The deformations that we consider are then actually equivalent to non-abelian T-duality on the Jordanian (super)algebras with no further deformation. See~\cite{Hoare:2016wsk,Borsato:2016pas,Borsato:2017qsx} for more details. 

Each of the Jordanian solutions that we have constructed are inequivalent and correspond to a different deformed supergravity background. Our results may therefore  offer a wide range of applications for deformations of the $AdS_5$/SYM holographic duality. See~\cite{vanTongeren:2015uha,vanTongeren:2016eeb} for some preliminary proposals of deformations of the dual gauge theory. See also~\cite{Araujo:2017jkb,Araujo:2017jap} for later works. In this paper,  we were primarily concerned with the algebraic classification of (unimodular) Jordanian solutions. For the purpose of deformed holography, one may in a first stage also analyse if the dilaton is well-behaved for the simpler unimodular $R$-matrices of table \ref{tab:rank-2-siso-full}. In this table, the case $\bar{R}_1$ ($a=0$, $b=-1/2$, $q^+_{2,22}=0$) is the only one with 12 residual superisometries. It is in fact this example (up to inner automorphisms) which was first constructed in \cite{vanTongeren:2019dlq} (and indeed has a well-behaved dilaton) and of which the semi-classical spectral problem of the deformed sigma-model was later analysed in \cite{Borsato:2022drc} by means of algebraic curve techniques.

Instead of considering the deformed backgrounds that preserve the maximal number of superisometries, one could actually study  deformations that break supersymmetry completely. This direction is particularly interesting because integrability may prove to be a new guiding principle to understand non-supersymmetric backgrounds. Backgrounds with low or no supersymmetry will arise also from the higher-rank solutions classified in this paper, and it would be interesting to carry out also an explicit analysis of their preserved superisometries.

It is pressing to understand how to extend the methods of quantum integrability in order to obtain the exact spectrum of the Jordanian-deformed string. The main obstacle in doing so is the breaking of the BMN isometries in the presence of any of the Jordanian deformations classified in this paper. This implies that we should not expect the Jordanian deformations of the quantum integrable model to be implemented in a simple way, in particular they will not be deformations of the undeformed S-matrix of Beisert~\cite{Beisert:2005tm}. It is possible that the findings of~\cite{Borsato:2022drc} in the classical setup will be useful in order to clarify these issues, and we hope to address this problem in the near future.

As we have found, the family of Jordanian deformations of $AdS_5\times S^5$ is quite large. This is a confirmation of a large landscape of integrable deformations of the string sigma-model, that undoubtedly extends beyond the Jordanian class.

\section*{Acknowledgements}
We thank Leander Wyss for useful discussions and collaboration during the initial stages of this project. 
We are supported by the fellowship of ``la Caixa Foundation'' (ID 100010434) with code LCF/BQ/PI19/11690019, by AEI-Spain (under project PID2020-114157GB-I00 and Unidad de Excelencia Mar\'\i a de Maetzu MDM-2016-0692),  Xunta de Galicia (Centro singular de investigaci\'on de Galicia accreditation 2019-2022, and project ED431C-2021/14), and by the European Union FEDER. This research was supported in part by the National Science Foundation under Grant No. NSF PHY-1748958.


\appendix

\section{Homogeneous Yang-Baxter deformations}\label{app:conv}

In this appendix, we collect facts regarding solutions of the classical Yang-Baxter equation and our related conventions.
Let us consider a generic Lie superalgebra $\alg g$ (the truncation of these facts to bosonic Lie subalgebras is straightforward). We are after ``antisymmetric'' constant solutions of the classical Yang-Baxter equation on $\alg g$. That means that we want to construct a linear operator 
\begin{equation}
    R:\ \alg g\to \alg g,
\end{equation}
that is antisymmetric with respect to a bilinear form on $\alg g$. For simplicity here we assume that the latter is implemented by taking the (super)trace in a matrix realisation of $\alg g$, so that antisymmetry reads like
\begin{equation}\label{eq:antisymmR}
    \STr(Rx\ y)=-\STr(x\ Ry),\qquad
    \forall x,y\in \alg g.
\end{equation}
Moreover, we demand that it solves the classical Yang-Baxter equation which reads
\begin{equation} \label{eq:CYBE}
    [[Rx,Ry]]-R([[Rx,y]]+[[x,Ry]])=0,\qquad \forall x,y\in \alg g,
\end{equation}
where $[[,]]$ denotes the graded commutator on the superalgebra (i.e.~it is the anticommutator when the two elements are odd and the commutator otherwise).

Let us introduce a basis $\T_I$ for $\alg g$ to identify the structure constants as $[[T_I,T_J]]=f_{IJ}{}^KT_K$. The linear operator $R$ can be thought of as a matrix $R_J{}^J$ because $R\T_I=R_I{}^J\T_J$. Moreover, after denoting by $K_{IJ}=\STr(\T_I\T_J)$ the metric on $\alg g$ and $K^{IJ}$ its inverse, antisymmetry just corresponds to the statement that $R^{IJ}=K^{IK}R_K{}^J$ is antisymmetric in the $I,J$ indices if the indices correspond to even generators of $\alg g$, while $R^{IJ}$ is symmetric if the indices correspond to odd generators. This difference is due to an extra sign coming from the supertrace.

We can also map $R$ to an element $r$ of the 2-fold wedge product of $\alg g$ by
\begin{equation} \label{eq:r-to-R}
    r=-\tfrac12 R^{IJ}\T_I\wedge \T_J,
\end{equation}
where we use the graded wedge product
\begin{equation}
    x\wedge y=x\otimes y-(-1)^{\text{deg}(x)*\text{deg}(y)}\, y\otimes x.
\end{equation}
In this definition we are using the function deg that gives the degree of the superalgebra element, i.e.~it is 0 on even elements and 1 on odd ones. This implies that $\wedge$ is symmetric if both $x$ and $y$ are odd, otherwise it is antisymmetric.

The above algebraic ingredients can be used to construct integrable deformations of 2-dimensional sigma models. In the case of deformations of semisymmetric supercoset sigma-models we assume the existence of a $\mathbb{Z}_4$-grading of $\alg g$ such that $\alg g=\oplus_{i=0}^3\alg g^{(i)}$ and $[[\alg g^{(i)},\alg g^{(j)}]]\subset \alg g^{(i+j\text{ mod }4)}$. Then the action of the deformed sigma model is \cite{Delduc:2013qra}
\begin{equation} \label{eq:hyb-ssssm}
    S=-\frac{\sqrt \lambda}{8\pi}\int d\tau d\sigma\, (\sqrt{|h|}h^{mn}-\varepsilon^{mn})\, \STr\left(J_m\, \hat d\frac{1}{1-\eta R_g\hat d}J_n\right).
\end{equation}
We have introduced a worldsheet metric $h_{mn}$ and the antisymmetric tensor $\varepsilon^{\tau\sigma}=-\varepsilon^{\sigma\tau}=-1$. We also have the Maurer-Cartan form $J=g^{-1}dg$ with $g\in G$ and $\alg g=\mathrm{Lie}(G)$, and $\hat d=\tfrac12 P^{(1)}+P^{(2)}-\tfrac12 P^{(3)}$ with $P^{(i)}$ the projectors on $\mathfrak{g}^{(i)}$. The shorthand notation $R_g$ means $R_g=\AD_g^{-1}R\AD_g$ where $\AD_g x=gxg^{-1}$ and it is multiplied by the deformation parameter $\eta\in\mathbb{R}$.

It is now very simple to obtain the truncation to symmetric bosonic cosets. In that case, one assumes a $\mathbb Z_2$-grading and  replaces the supertrace by the trace as well as $\hat d=P^{(2)}$. To reduce even further to deformations of the Principal Chiral Model, we only have to take $\hat d=1$.

Importantly, the previous algebraic ingredients can be used also to generate deformations of supergravity backgrounds that do not necessarily correspond to integrable sigma-models. In order to make sure that the deformation still solves the type II supergravity equations, the $R$-matrix must satisfy an additional linear constraint called the ``unimodularity condition'' \cite{Borsato:2016ose}
\begin{equation} \label{eq:uni-cond}
    0=K^{IJ}[[\T_I,R\T_J]]=R^{IJ}f_{IJ}{}^K\T_K.
\end{equation}

To conclude, let us remark that we are interested only in real deformations of the supergravity background. Demanding the reality of the sigma-model action we find that the $R$-matrix gives rise to a real deformation if $R^{IJ}$ is anti-hermitian $(R^{JI})^*=-R^{IJ}$. This result follows from assuming the reality condition $T_I^\dagger+HT_IH^{-1}=0$ for elements of $\alg g$. See appendix~\ref{app:psu224} for the case of $\alg{psu}(2,2|4)$. As remarked above, on the one hand, $R^{IJ}$ is antisymmetric if the indices correspond to even generators of $\alg g$ and then the entries of $R^{IJ}$ must be real. On the other hand, $R^{IJ}$ is symmetric if the indices correspond to odd generators of $\alg g$ and then the entries of $R^{IJ}$ must be imaginary.

\section{The $\alg{psu}(2,2|4)$ superalgebra}\label{app:psu224}
Here we collect our conventions for the $\mathcal N=4$ superconformal algebra, and we provide an explicit matrix realisation of $\mathfrak{su}(2,2|4)$ that is useful for explicit calculations. 
Useful reviews are for example~\cite{Minahan:2010js} and~\cite{Arutyunov:2009ga}.

\subsection*{Indices conventions}
We will use $\mu,\nu=0,\ldots,3$ for indices in the 4-dimensional spacetime and we will take the Minkowski metric to be $\eta_{\mu\nu}=\text{diag}(-1,+1,+1,+1,)$. Knowing that the Lorentz algebra can be rewritten as $\mathfrak{so}(1,3)\sim \mathfrak{sl}(2,\mathbb R)_L\oplus \mathfrak{sl}(2,\mathbb R)_R$, we will use $\alpha,\beta=1,2$ for spinor indices of $\mathfrak{sl}(2,\mathbb R)_L$ and $\dot\alpha,\dot\beta=1,2$ for spinor indices of $\mathfrak{sl}(2,\mathbb R)_R$. Finally, we will  use $a,b=1,\ldots,4$ for spinor indices of $SO(6)$, and $A,B=1,\ldots 6$ for fundamental indices of $SO(6)$. 

\subsection*{The conformal algebra $\alg{so}(2,4)$}

The Lorentz algebra is spanned by $J_{\mu\nu}$ satisfying
\begin{equation}
    [J_{\mu\nu},J_{\rho\sigma}]=\eta_{\mu\rho}J_{\nu\sigma}-\eta_{\nu\rho}J_{\mu\sigma}+\eta_{\nu\sigma}J_{\mu\rho}-\eta_{\mu\sigma}J_{\nu\rho}.
\end{equation}
With the addition of $p_\mu$ they form the Poincar\'e algebra
\begin{equation}
    [J_{\mu\nu},p_\rho]=\eta_{\mu\rho}p_\nu-\eta_{\nu\rho}p_\mu.
\end{equation}
Adding the dilatation generator $D$ and the special conformal generators $k_\mu$ we obtain the full conformal algebra, whose remaining commutation relations are
\begin{equation}
    [D,p_\mu]=p_\mu,\qquad
    [D,k_\mu]=-k_\mu,\qquad
    [p_\mu,k_\nu]=-2\eta_{\mu\nu}D+2J_{\mu\nu},\qquad
    [J_{\mu\nu},k_\rho]=\eta_{\mu\rho}k_\nu-\eta_{\nu\rho}k_\mu.
\end{equation}
All other commutation relations are trivial
\begin{equation}
    [p_\mu,p_\nu]=[k_\mu,k_\nu]=[D,J_{\mu\nu}]=0.
\end{equation}

\subsection*{Supercharges and $R$-symmetry}
At this point we introduce supercharges $Q_{\alpha a},\overline Q^{\dot\alpha a},S_\alpha{}^a,\overline S^{\dot \alpha}{}_a$. 
The generators $J_{\mu\nu},p_\mu,Q_{\alpha a},\overline Q^{\dot\alpha a}$ span the ($\mathcal N=4$) super-Poincar\'e algebra. A lower (resp.~upper) index $\alpha$ means that they transform in the $\mathbf 2$ (resp.~$\overline{\mathbf 2}$) representation of $\mathfrak{sl}(2,\mathbb R)_L$, and similarly for dotted indices of  $\mathfrak{sl}(2,\mathbb R)_R$. To be more explicit, let us define the antisymmetric tensor $\epsilon_{12}=-\epsilon_{21}=-\epsilon^{12}=\epsilon^{21}=1$ such that $\epsilon^{\alpha\gamma}\epsilon_{\gamma\beta}=\delta^\alpha_\beta$. This is used to raise and lower indices as $\psi^\alpha=\epsilon^{\alpha\beta}\psi_\beta,\psi_\alpha=\epsilon_{\alpha\beta}\psi^\beta$.
We then define $(\sigma^\mu)_{\alpha\dot \alpha}$ and  $(\overline\sigma^\mu)^{\dot \alpha\alpha}$ as
\begin{equation}
    \sigma^\mu=-i\, (\mathbf 1,\sigma^j),\qquad
    \overline\sigma^\mu=-i\, (\mathbf 1,-\sigma^j),
\end{equation}
where $\sigma^j$ are the Pauli matrices. The matrices $\sigma^\mu, \overline{\sigma}^\mu$ appear in the Weyl representation of the 4-dimensional gamma-matrices
\begin{equation}\label{eq:Weyl-gamma}
    \gamma^\mu=\left(\begin{array}{cc}
        0 & (\sigma^\mu)_{\alpha\dot \alpha} \\
       (\overline\sigma^\mu)^{\dot \alpha\alpha}  & 0
    \end{array}
    \right),
\end{equation}
that satisfy $\{\gamma_\mu,\gamma_\nu\}=2\eta_{\mu\nu}\mathbf 1$.
At this point we define
\begin{equation}
    (\sigma^{\mu\nu})_\alpha{}^\beta=\tfrac14 (\sigma^\mu\overline\sigma^\nu-\sigma^\nu\overline\sigma^\mu)_\alpha{}^\beta,\qquad
    (\overline\sigma^{\mu\nu})^{\dot\alpha}{}_{\dot\beta}=\tfrac14 (\overline\sigma^\mu\sigma^\nu-\overline\sigma^\nu\sigma^\mu)^{\dot\alpha}{}_{\dot\beta},
\end{equation}
so that we can write the commutators of the supercharges with the Lorentz generators
\begin{equation}
    \begin{aligned}
        &[J_{\mu\nu},Q_{\alpha a}]=(\sigma_{\mu\nu})_\alpha{}^\beta Q_{\beta a},\qquad
        &&[J_{\mu\nu},\overline Q^{\dot\alpha a}]=(\overline\sigma_{\mu\nu})^{\dot\alpha}{}_{\dot\beta} \overline Q^{\dot\beta a},\\
        &[J_{\mu\nu},S_{\alpha}{}^a]=(\sigma_{\mu\nu})_\alpha{}^\beta S_{\beta}{}^a,\qquad
        &&[J_{\mu\nu},\overline S^{\dot\alpha}{}_a]=(\overline\sigma_{\mu\nu})^{\dot\alpha}{}_{\dot\beta} \overline S^{\dot\beta}{}_a.
    \end{aligned}
\end{equation}
We have the following trivial commutation relations
\begin{equation}
    [p_\mu,Q_{\alpha a}]=[p_\mu,\overline Q^{\dot\alpha a}]=[k_\mu,S_{\alpha}{}^a]=[k_\mu,\overline S^{\dot\alpha}{}_a]=0,
\end{equation}
and the commutation relations with the dilatation generator
\begin{equation} \label{eq:comm-D-supercharges}
    \begin{aligned}
& [D,Q_{\alpha a}]=\tfrac12 Q_{\alpha a},\qquad 
&& [D,S_{\alpha}{}^a]=-\tfrac12 S_{\alpha}{}^a,\\
& [D,\overline Q^{\dot\alpha a}]=\tfrac12 \overline Q^{\dot\alpha a},
&&[D,\overline S^{\dot\alpha}{}_a]=-\tfrac12 \overline S^{\dot\alpha}{}_a.        
    \end{aligned}
\end{equation}
The commutators relating the $Q$ and $S$ supercharges are
\begin{equation}
    \begin{aligned}
    &[k^\mu,Q_{\alpha a}]=+i \sigma^\mu_{\alpha\dot\alpha}\, \overline S^{\dot\alpha}{}_a,\qquad
    &&[k_\mu,\overline Q^{\dot\alpha a}]= -i\overline\sigma_\mu^{\dot\alpha\alpha}\,  S_{\alpha}{}^a,\\
        &[p^\mu,S_{\alpha}{}^a]=-i \sigma^\mu_{\alpha\dot\alpha}\, \overline Q^{\dot\alpha a},\qquad
    &&[p_\mu,\overline S^{\dot\alpha}{}_a]=+i \overline\sigma_\mu^{\dot\alpha\alpha}\,  Q_{\alpha a}.
    \end{aligned}
\end{equation}
The $\mathcal N=4$ superconformal algebra has an $SU(4)\sim SO(6)$ $\mathcal R$-symmetry, under which the supercharges transform in the $\mathbf 4$ or $\overline{\mathbf 4}$ representations (respectively for upper or lower indices $a,b,=1,\ldots, 4$). We denote the $\mathcal R$-symmetry generators as $\mathcal R_{AB}$ with $\mathcal R_{AB}=-\mathcal R_{BA}$ and $A,B=1,\dots,6$. They satisfy the commutation relations
\begin{equation} \label{eq:comm-R}
    [\mathcal R_{AB},\mathcal R_{CD}]=\delta_{AC}\mathcal R_{BD}-\delta_{BC}\mathcal R_{AD}+\delta_{BD}\mathcal R_{AC}-\delta_{AD}\mathcal R_{BC},
\end{equation}
and they commute with all generators of the conformal algebra. The action of the $\mathcal R$-symmetry generators on the supercharges yields
\begin{equation}\label{eq:comm-R-QS}
\begin{aligned}
    &[\mathcal R_{AB},Q_{\alpha a}]=\tfrac12 (\rho_{AB})_a{}^b\, Q_{\alpha b},\qquad
    &&[\mathcal R_{AB},\overline Q^{\dot \alpha a}]=-\tfrac12 (\rho_{AB})_b{}^a\, \overline Q^{\dot \alpha b}\\
    &[\mathcal R_{AB},\overline S^{\dot \alpha}{}_{a}]=\tfrac12 (\rho_{AB})_a{}^b\, \overline S^{\dot \alpha}{}_b,\qquad
    &&[\mathcal R_{AB},S_\alpha{}^a]=-\tfrac12 (\rho_{AB})_b{}^a\, S_\alpha{}^b.
\end{aligned}
\end{equation}
Indices $A,B$ will be raised and lowered with the Kronecker delta.
Simple anticommutators are
\begin{equation}
    \{Q_{\alpha a},\overline Q_{\dot\alpha}{}^{ b}\}= \delta_a^b\, \sigma^\mu_{\alpha\dot\alpha}\,  p_\mu,\qquad\qquad
        \{S_{\alpha}{}^a,\overline S_{\dot\alpha b}\}=-  \delta^a_b\,  \sigma^\mu_{\alpha\dot\alpha}\,k_\mu.
\end{equation}
The trivial mixed anticommutators are
\begin{equation}
    \{Q_{\alpha a},\overline S^{\dot \beta}{}_b\}=0,\qquad \{\overline Q^{\dot \alpha a},S_\beta{}^b\}=0,
\end{equation}
while the remaining non-trivial mixed anticommutators
\begin{equation}
\begin{aligned}
        &\{Q_{\alpha a},S_\beta{}^b\}=
    \tfrac{i}{2}\, \epsilon_{\alpha\beta}\,  (\rho^{AB})_a{}^b\, \mathcal R_{AB}
    +i\,  \delta_a^b\,  \sigma^{\mu\nu}_{\alpha\beta}\, J_{\mu\nu}
    +i\, \epsilon_{\alpha\beta}\, \delta_a^b\,  D +\tfrac{i}{2}\,  \epsilon_{\alpha\beta}\, \delta_a^b\, \mathbf 1,\\
    &\{\overline Q^{\dot \alpha a},\overline S^{\dot\beta}{}_b\}=
    \tfrac{i}{2}\, \epsilon^{\dot\alpha\dot\beta}\,  (\rho^{AB})_b{}^a\, \mathcal R_{AB}
    -i\,  \delta^a_b\, \overline \sigma_{\mu\nu}^{\dot\alpha\dot\beta}\, J^{\mu\nu}
    -i\, \epsilon^{\dot\alpha\dot \beta}\,  \delta^a_b\,  D +\tfrac{i}{2}\,  \epsilon^{\dot\alpha\dot\beta}\, \delta^a_b\, \mathbf 1.
\end{aligned}
\end{equation}
The relations that we are writing here actually correspond to the $\mathfrak{su}(2,2|4)$ superalgebra.  To obtain the relations of $\mathfrak{psu}(2,2|4)$ (which is isomorphic to the $\mathcal N=4$ superconformal algebra) one has to project out the identity operator. 

\subsection*{Matrix realisation}
In the anticommutators above we included the terms proportional to the identity operator because we want to give an explicit matrix realisation of the superalgebra, and $\mathfrak{su}(2,2|4)$ admits one while $\mathfrak{psu}(2,2|4)$ does not.
To obtain the matrix realisation we start from the above definition~\eqref{eq:Weyl-gamma} of the gamma matrices, which is also equivalent to
\begin{equation}
    \begin{aligned}
            \gamma^0=-i \sigma^1\otimes \mathbf 1_2,\qquad \gamma^1= \sigma^2\otimes \sigma^1,\qquad
            \gamma^2= \sigma^2\otimes \sigma^2,\qquad \gamma^3= \sigma^2\otimes \sigma^3,
    \end{aligned}
\end{equation}
and we supplement them with 
\begin{equation}
    \gamma^4=-\sigma^3\otimes \mathbf 1_2,
\end{equation}
to obtain gamma matrices in 5 dimensions. In general, we define $\gamma_{mn}=\tfrac12[\gamma_m,\gamma_n]$ with indices $m,n=0,\ldots, 5$ and
\begin{equation}
    \tilde\gamma_i=\gamma_i,\quad i=1,\ldots,4,\qquad 
    \tilde \gamma_5=i\, \gamma_0,
\end{equation}
which are gamma-matrices in 5 \emph{Euclidean} dimensions. We use them to define the matrices $\rho_{AB}$ which are antisymmetric in the indices $A,B=1,\dots,6$ as
\begin{equation}
    \rho_{AB}=\tilde\gamma_{AB},\quad A,B=1,\ldots,5,\qquad \rho_{A6}=-i\, \tilde \gamma_A.
\end{equation}
Finally, we have everything we need to construct a matrix realisation of $\mathfrak{su}(2,2|4)$ in terms of $8\times 8$ matrices. For the generators of the conformal algebra we take
\begin{equation}
\begin{aligned}
            J_{\mu\nu}& =\left(\begin{array}{cc}
    -\tfrac12 \gamma_{\mu\nu}     & \mathbf 0_4  \\
    
        \mathbf 0_4 & \mathbf 0_4
    \end{array}\right),\qquad 
   && p_{\mu} =\left(\begin{array}{cc}
    -\tfrac12 (\gamma_{\mu 4}+\gamma_\mu  )   & \mathbf 0_4 \\
        \mathbf 0_4 & \mathbf 0_4
    \end{array}\right),\\ 
    D &=\left(\begin{array}{cc}
    -\tfrac12 \gamma_4     & \mathbf 0_4  \\
        \mathbf 0_4 & \mathbf 0_4 
    \end{array}\right),\qquad 
   && k_{\mu} =\left(\begin{array}{cc}
    -\tfrac12 (\gamma_{\mu 4}-\gamma_\mu  )   & \mathbf 0_4  \\
        \mathbf 0_4 & \mathbf 0_4
    \end{array}\right).
\end{aligned}
\end{equation}
Similarly, for the $\mathcal R$-symmetry generators
\begin{equation}
            \mathcal R_{AB} =\left(\begin{array}{cc}
 \mathbf 0_4   & \mathbf 0_4  \\
    
        \mathbf 0_4 &    -\tfrac12 \rho_{AB}  
    \end{array}\right).
\end{equation}
To conclude, the supercharges are realised as
\begin{equation}
\begin{aligned}
            &Q^\alpha{}_a=\sqrt{2}\left(\begin{array}{cc}
   \mathbf 0_4   & E_{\alpha, a} \\
        \mathbf 0_4 & \mathbf 0_4
    \end{array}\right),\qquad 
        &&\overline Q^{\dot\alpha a}=-\sqrt{2}\left(\begin{array}{cc}
   \mathbf 0_4   &    \mathbf 0_4 \\
    E_{a,\dot\alpha+2}  & \mathbf 0_4
    \end{array}\right),\\
    &S_\alpha{}^a=i\sqrt{2}\left(\begin{array}{cc}
   \mathbf 0_4   & \mathbf 0_4 \\
         E_{a,\alpha} & \mathbf 0_4
    \end{array}\right),\qquad 
        &&\overline S_{\dot\alpha a}=i\sqrt{2}\left(\begin{array}{cc}
   \mathbf 0_4   &  E_{\dot\alpha+2,a} \\
        \mathbf 0_4 & \mathbf 0_4
    \end{array}\right),
\end{aligned}    
\end{equation}
where $E_{a,b}$ are the $4\times 4$ unit matrices with zeros everywhere, except  1 at position $a,b$.

\subsection*{Reality condition on the superalgebra}
To write down the reality condition, let us define
\begin{equation}
    H=\left(\begin{array}{cc}
   -i\gamma_0  & \mathbf 0_4 \\
        \mathbf 0_4  & \mathbf 1_4
    \end{array}\right).
\end{equation}
With this choice $H^\dagger = H$,  where $\dagger$ denotes conjugate-transpose.
For all bosonic generators $X$ (i.e.~from the conformal  or the $\mathcal R$-symmetry algebra) the reality condition is satisfied as $X^\dagger +HXH^{-1}=0$. For supercharges, instead, the dagger relates the barred and unbarred supercharges in the following way
\begin{equation}\label{eq:reality-QS}
    (Q_{\alpha a})^\dagger +H\overline Q_\alpha{}^aH^{-1}=0,\qquad \qquad 
    (S_{\alpha}{}^a)^\dagger +H\overline S_{\alpha a}H^{-1}=0.
\end{equation}
This means that we are using a complex basis. A generic  element $M$ of $\mathfrak{psu}(2,2|4)$, however, is required to satisfy  simply $M^\dag + H M H^{-1}=0$.

\subsection*{$\mathbb Z_4$ automorphism}
The $\mathcal N=4$ superconformal algebra admits a $\mathbb Z_4$ automorphism. Let us define the matrix \begin{equation}
    K_{ab}=-i (\mathbf 1_2 \otimes \sigma_2)_{ab},
\end{equation}
and let us denote by $K^{ab}$ its inverse, so that $K^{ac}K_{cb}=\delta^a_b$. We will use $K$ to raise and lower $a,b$ indices, with the same conventions as for the Lorentz indices, namely $V_a=K_{ab}V^b,\ V^a=K^{ab}V_b$. We also define
\begin{equation}
    \mathcal K=\mathbf 1_2\otimes K,
\end{equation}
which we use for the definition of the $\mathbb Z_4$ automorphism as
\begin{equation}
    \Omega(X)=-\mathcal K\, X^{st}\, \mathcal K^{-1},
\end{equation}
where $st$ denotes supertransposition. The  $\mathbb Z_4$ automorphism induces the decomposition $\alg g=\oplus_{i=0}^3\alg g^{(i)}$ of the superalgebra and one can construct the projectors $P^{(i)}$ on each of these subspaces. Then we find the following decomposition
\begin{equation}
\begin{aligned}
     &  p_\mu+k_\mu,\ J_{\mu\nu},\ \mathcal R_{\bar A\bar B}\in  \mathfrak g^{(0)}\qquad && Q_{\alpha a}+S_{\alpha a},\ \overline{Q}^{\dot \alpha a}- \overline{S}^{\dot \alpha a}\in  \mathfrak g^{(1)},\\
   & p_\mu-k_\mu,\ D,\ \mathcal R_{\bar A 6}\ \in \mathfrak g^{(2)},\qquad&&Q_{\alpha a}- S_{\alpha a},\ \overline{Q}^{\dot \alpha a}+ \overline{S}^{\dot \alpha a}\in  \mathfrak g^{(3)},
\end{aligned}
\end{equation}
where above $\bar A,\bar B=1,\ldots,5$.
We remind that indices are raised and lowered with $\epsilon$ and $K$.
As reviewed above, in the construction of the supercoset action and its deformations, one introduces a particular combination of the projectors which is $\hat d=P^{(1)}+2P^{(2)}-P^{(3)}$. The action of its transpose $\hat d^T=-P^{(1)}+2P^{(2)}+P^{(3)}$ on the supercharges is
\begin{equation}
\hat     d^T(Q_{\alpha a})=-S_{\alpha a},\qquad 
    \hat d^T(S_{\alpha a})=-Q_{\alpha a},\qquad 
    \hat d^T(\overline Q^{\dot\alpha a})=\overline S^{\dot\alpha a},\qquad 
    \hat d^T(\overline S^{\dot\alpha a})=\overline Q^{\dot\alpha a}.
\end{equation}

\subsection*{Supertrace relations}
The non-vanishing relations involving the supertrace are
\begin{equation}
    \begin{aligned}
     &  \STr(p_\mu k_\nu)=2\eta_{\mu\nu},\qquad 
     &&  \STr(J_{\mu\nu}J_{\rho\sigma})=-(\eta_{\mu\rho}\eta_{\nu\sigma}-\eta_{\nu\rho}\eta_{\mu\sigma}),\\ 
   &    \STr(DD)=1,\qquad 
    &&   \STr(\mathcal R_{AB}\mathcal R_{CD})=\delta_{AC}\delta_{BD}-\delta_{BC}\delta_{AD}, \\
    & \STr(Q_{\alpha a}S_\beta{}^b)=2i\, \epsilon_{\alpha\beta}\delta_a^b,\qquad 
    && \STr(\overline Q^{\dot\alpha a}\overline S^{\dot\beta}{}_b)=-2i\, \epsilon^{\dot\alpha\dot\beta}\delta^a_b.
    \end{aligned}
\end{equation}

\bibliographystyle{nb}
\bibliography{biblio}{}

\begin{thebibliography}{10}
\ifx\href\asklfhas\newcommand{\href}[2]{#2}\fi
\ifx\arxivref\asklfhas\newcommand{\arxivref}[2]{\href{http://arxiv.org/abs/#1}{#2}}\fi
\ifx\doiref\asklfhas\newcommand{\doiref}[2]{\href{http://dx.doi.org/#1}{#2}}\fi
\raggedright
\small
\parskip 0pt

\bibitem{Lunin:2005jy}
O.~Lunin and J.~M.~Maldacena,
\textit{``{Deforming field theories with $U(1) \times U(1)$ global symmetry and
  their gravity duals}''},
\textsf{\doiref{10.1088/1126-6708/2005/05/033}{JHEP~0505,~033~(2005)}},
\texttt{\arxivref{hep-th/0502086}{hep-th/0502086}}.

\bibitem{Frolov:2005dj}
S.~Frolov,
\textit{``{Lax pair for strings in Lunin-Maldacena background}''},
\textsf{\doiref{10.1088/1126-6708/2005/05/069}{JHEP~0505,~069~(2005)}},
\texttt{\arxivref{hep-th/0503201}{hep-th/0503201}}.

\bibitem{Frolov:2005ty}
S.~A.~Frolov, R.~Roiban and A.~A.~Tseytlin,
\textit{``{Gauge-string duality for superconformal deformations of N=4 super
  Yang-Mills theory}''},
\textsf{\doiref{10.1088/1126-6708/2005/07/045}{JHEP~0507,~045~(2005)}},
\texttt{\arxivref{hep-th/0503192}{hep-th/0503192}}.

\bibitem{Klimcik:2002zj}
C.~Klimcik,
\textit{``{Yang-Baxter sigma models and dS/AdS T duality}''},
\textsf{\doiref{10.1088/1126-6708/2002/12/051}{JHEP~0212,~051~(2002)}},
\texttt{\arxivref{hep-th/0210095}{hep-th/0210095}}.

\bibitem{Klimcik:2008eq}
C.~Klimcik,
\textit{``{On integrability of the Yang-Baxter sigma-model}''},
\textsf{\doiref{10.1063/1.3116242}{J.Math.Phys.~50,~043508~(2009)}},
\texttt{\arxivref{0802.3518}{arxiv:0802.3518}}.

\bibitem{Delduc:2013fga}
F.~Delduc, M.~Magro and B.~Vicedo,
\textit{``{On classical $q$-deformations of integrable sigma-models}''},
\textsf{\doiref{10.1007/JHEP11(2013)192}{JHEP~1311,~192~(2013)}},
\texttt{\arxivref{1308.3581}{arxiv:1308.3581}}.

\bibitem{Delduc:2013qra}
F.~Delduc, M.~Magro and B.~Vicedo,
\textit{``{An integrable deformation of the AdS$_5 \times$S$^5$ superstring
  action}''},
\textsf{\doiref{10.1103/PhysRevLett.112.051601}{Phys.Rev.Lett.~112,~051601~(2014)}},
\texttt{\arxivref{1309.5850}{arxiv:1309.5850}}.

\bibitem{Kawaguchi:2014qwa}
I.~Kawaguchi, T.~Matsumoto and K.~Yoshida,
\textit{``{Jordanian deformations of the $AdS_5 x S^5$ superstring}''},
\textsf{\doiref{10.1007/JHEP04(2014)153}{JHEP~1404,~153~(2014)}},
\texttt{\arxivref{1401.4855}{arxiv:1401.4855}}.

\bibitem{vanTongeren:2015soa}
S.~J.~van~Tongeren,
\textit{``{On classical Yang-Baxter based deformations of the AdS$_{5}$ ×
  S$^{5}$ superstring}''},
\textsf{\doiref{10.1007/JHEP06(2015)048}{JHEP~1506,~048~(2015)}},
\texttt{\arxivref{1504.05516}{arxiv:1504.05516}}.

\bibitem{Sfetsos:2013wia}
K.~Sfetsos,
\textit{``{Integrable interpolations: From exact CFTs to non-Abelian
  T-duals}''},
\textsf{\doiref{10.1016/j.nuclphysb.2014.01.004}{Nucl.Phys.~B880,~225~(2014)}},
\texttt{\arxivref{1312.4560}{arxiv:1312.4560}}.

\bibitem{Hollowood:2014rla}
T.~J.~Hollowood, J.~L.~Miramontes and D.~M.~Schmidtt,
\textit{``{Integrable Deformations of Strings on Symmetric Spaces}''},
\textsf{\doiref{10.1007/JHEP11(2014)009}{JHEP~1411,~009~(2014)}},
\texttt{\arxivref{1407.2840}{arxiv:1407.2840}}.

\bibitem{Hollowood:2014qma}
T.~J.~Hollowood, J.~L.~Miramontes and D.~M.~Schmidtt,
\textit{``{An Integrable Deformation of the AdS$_5 \times$S$^5$
  Superstring}''},
\textsf{\doiref{10.1088/1751-8113/47/49/495402}{J.Phys.~A47,~495402~(2014)}},
\texttt{\arxivref{1409.1538}{arxiv:1409.1538}}.

\bibitem{Thompson:2019ipl}
D.~C.~Thompson,
\textit{``{An Introduction to Generalised Dualities and their Applications to
  Holography and Integrability}''},
\textsf{\doiref{10.22323/1.347.0099}{PoS~CORFU2018,~099~(2019)}},
\texttt{\arxivref{1904.11561}{arxiv:1904.11561}}.

\bibitem{Hoare:2021dix}
B.~Hoare,
\textit{``{Integrable deformations of sigma models}''},
\textsf{\doiref{10.1088/1751-8121/ac4a1e}{J.~Phys.~A~55,~093001~(2022)}},
\texttt{\arxivref{2109.14284}{arxiv:2109.14284}}.

\bibitem{Borsato:2016ose}
R.~Borsato and L.~Wulff,
\textit{``{Target space supergeometry of $\eta$ and $\lambda$-deformed
  strings}''},
\textsf{\doiref{10.1007/JHEP10(2016)045}{JHEP~1610,~045~(2016)}},
\texttt{\arxivref{1608.03570}{arxiv:1608.03570}}.

\bibitem{Hoare:2018ngg}
B.~Hoare and F.~K.~Seibold,
\textit{``{Supergravity backgrounds of the $\eta$-deformed AdS$_2 \times S^2
  \times T^6 $ and AdS$_5 \times S^5$ superstrings}''},
\textsf{\doiref{10.1007/JHEP01(2019)125}{JHEP~1901,~125~(2019)}},
\texttt{\arxivref{1811.07841}{arxiv:1811.07841}}.

\bibitem{Sfetsos:2014cea}
K.~Sfetsos and D.~C.~Thompson,
\textit{``{Spacetimes for $\lambda$-deformations}''},
\textsf{\doiref{10.1007/JHEP12(2014)164}{JHEP~1412,~164~(2014)}},
\texttt{\arxivref{1410.1886}{arxiv:1410.1886}}.

\bibitem{Appadu:2015nfa}
C.~Appadu and T.~J.~Hollowood,
\textit{``{Beta function of k deformed AdS$_{5}$ \texttimes{} S$^{5}$ string
  theory}''},
\textsf{\doiref{10.1007/JHEP11(2015)095}{JHEP~1511,~095~(2015)}},
\texttt{\arxivref{1507.05420}{arxiv:1507.05420}}.

\bibitem{Borsato:2021gma}
R.~Borsato and S.~Driezen,
\textit{``{Supergravity solution-generating techniques and canonical
  transformations of $\sigma$-models from $O(D,D)$}''},
\textsf{\doiref{10.1007/JHEP05(2021)180}{JHEP~2105,~180~(2021)}},
\texttt{\arxivref{2102.04498}{arxiv:2102.04498}}.

\bibitem{Borsato:2021vfy}
R.~Borsato, S.~Driezen and F.~Hassler,
\textit{``{An algebraic classification of solution generating techniques}''},
\textsf{\doiref{10.1016/j.physletb.2021.136771}{Phys.~Lett.~B~823,~136771~(2021)}},
\texttt{\arxivref{2109.06185}{arxiv:2109.06185}}.

\bibitem{Beisert:2010jr}
N.~Beisert et~al.,
\textit{``{Review of AdS/CFT Integrability: An Overview}''},
\textsf{\doiref{10.1007/s11005-011-0529-2}{Lett.~Math.~Phys.~99,~3~(2012)}},
\texttt{\arxivref{1012.3982}{arxiv:1012.3982}}.

\bibitem{Borsato:2017qsx}
R.~Borsato and L.~Wulff,
\textit{``{On non-abelian T-duality and deformations of supercoset string
  sigma-models}''},
\textsf{\doiref{10.1007/JHEP10(2017)024}{JHEP~1710,~024~(2017)}},
\texttt{\arxivref{1706.10169}{arxiv:1706.10169}}.

\bibitem{Borsato:2018idb}
R.~Borsato and L.~Wulff,
\textit{``{Non-abelian T-duality and Yang-Baxter deformations of Green-Schwarz
  strings}''},
\textsf{\doiref{10.1007/JHEP08(2018)027}{JHEP~1808,~027~(2018)}},
\texttt{\arxivref{1806.04083}{arxiv:1806.04083}}.

\bibitem{Arutyunov:2015mqj}
G.~Arutyunov, S.~Frolov, B.~Hoare, R.~Roiban and A.~A.~Tseytlin,
\textit{``{Scale invariance of the $\eta$-deformed $AdS_5\times S^5$
  superstring, T-duality and modified type II equations}''},
\textsf{\doiref{10.1016/j.nuclphysb.2015.12.012}{Nucl.~Phys.~B903,~262~(2016)}},
\texttt{\arxivref{1511.05795}{arxiv:1511.05795}}.

\bibitem{Wulff:2016tju}
L.~Wulff and A.~A.~Tseytlin,
\textit{``{Kappa-symmetry of superstring sigma model and generalized 10d
  supergravity equations}''},
\textsf{\doiref{10.1007/JHEP06(2016)174}{JHEP~1606,~174~(2016)}},
\texttt{\arxivref{1605.04884}{arxiv:1605.04884}}.

\bibitem{vanTongeren:2019dlq}
S.~J.~van~Tongeren,
\textit{``{Unimodular jordanian deformations of integrable superstrings}''},
\textsf{\doiref{10.21468/SciPostPhys.7.1.011}{SciPost~Phys.~7,~011~(2019)}},
\texttt{\arxivref{1904.08892}{arxiv:1904.08892}}.

\bibitem{Osten:2016dvf}
D.~Osten and S.~J.~van~Tongeren,
\textit{``{Abelian Yang–Baxter deformations and TsT transformations}''},
\textsf{\doiref{10.1016/j.nuclphysb.2016.12.007}{Nucl.~Phys.~B915,~184~(2017)}},
\texttt{\arxivref{1608.08504}{arxiv:1608.08504}}.

\bibitem{Alday:2005ww}
L.~F.~Alday, G.~Arutyunov and S.~Frolov,
\textit{``{Green-Schwarz strings in TsT-transformed backgrounds}''},
\textsf{\doiref{10.1088/1126-6708/2006/06/018}{JHEP~0606,~018~(2006)}},
\texttt{\arxivref{hep-th/0512253}{hep-th/0512253}}.

\bibitem{Beisert:2005if}
N.~Beisert and R.~Roiban,
\textit{``{Beauty and the twist: The Bethe ansatz for twisted N=4 SYM}''},
\textsf{\doiref{10.1088/1126-6708/2005/08/039}{JHEP~0508,~039~(2005)}},
\texttt{\arxivref{hep-th/0505187}{hep-th/0505187}}.

\bibitem{deLeeuw:2012hp}
M.~de~Leeuw and S.~J.~van~Tongeren,
\textit{``{The spectral problem for strings on twisted $AdS_5 \times S^5$}''},
\textsf{\doiref{10.1016/j.nuclphysb.2012.03.004}{Nucl.~Phys.~B860,~339~(2012)}},
\texttt{\arxivref{1201.1451}{arxiv:1201.1451}}.

\bibitem{Kazakov:2018ugh}
V.~Kazakov,
\textit{``{Quantum Spectral Curve of $\gamma$-twisted ${\cal N}=4$ SYM theory
  and fishnet CFT}''},
\texttt{\arxivref{1802.02160}{arxiv:1802.02160}},
in: \textit{``Ludwig Faddeev Memorial Volume''},
ed.: M.-L.~Ge, A.~J.~Niemi, K.~K.~Phua and L.~A.~Takhtajan,
293--342p.

\bibitem{Meier:2023kzt}
T.~Meier and S.~J.~van~Tongeren,
\textit{``{Quadratic twist-noncommutative gauge theory}''},
\texttt{\arxivref{2301.08757}{arxiv:2301.08757}}.

\bibitem{Borsato:2021fuy}
R.~Borsato, S.~Driezen and J.~L.~Miramontes,
\textit{``{Homogeneous Yang-Baxter deformations as undeformed yet twisted
  models}''},
\textsf{\doiref{10.1007/JHEP04(2022)053}{JHEP~2204,~053~(2022)}},
\texttt{\arxivref{2112.12025}{arxiv:2112.12025}}.

\bibitem{Matsumoto:2015jja}
T.~Matsumoto and K.~Yoshida,
\textit{``{Yang–Baxter sigma models based on the CYBE}''},
\textsf{\doiref{10.1016/j.nuclphysb.2015.02.009}{Nucl.~Phys.~B893,~287~(2015)}},
\texttt{\arxivref{1501.03665}{arxiv:1501.03665}}.

\bibitem{Vicedo:2015pna}
B.~Vicedo,
\textit{``{Deformed integrable \ensuremath{\sigma}-models, classical R-matrices
  and classical exchange algebra on Drinfel\textquoteright{}d doubles}''},
\textsf{\doiref{10.1088/1751-8113/48/35/355203}{J.~Phys.~A~48,~355203~(2015)}},
\texttt{\arxivref{1504.06303}{arxiv:1504.06303}}.

\bibitem{vanTongeren:2018vpb}
S.~J.~Van~Tongeren,
\textit{``{On Yang--Baxter models, twist operators, and boundary
  conditions}''},
\textsf{\doiref{10.1088/1751-8121/aac8eb}{J.~Phys.~A~51,~305401~(2018)}},
\texttt{\arxivref{1804.05680}{arxiv:1804.05680}}.

\bibitem{Borsato:2022drc}
R.~Borsato, S.~Driezen, J.~M.~Nieto~Garc\'\i{}a and L.~Wyss,
\textit{``{Semiclassical spectrum of a Jordanian deformation of
  AdS5\texttimes{}S5}''},
\textsf{\doiref{10.1103/PhysRevD.106.066015}{Phys.~Rev.~D~106,~066015~(2022)}},
\texttt{\arxivref{2207.14748}{arxiv:2207.14748}}.

\bibitem{Borsato:2018spz}
R.~Borsato and L.~Wulff,
\textit{``{Marginal deformations of WZW models and the classical Yang-Baxter
  equation}''},
\textsf{\doiref{10.1088/1751-8121/ab1b9c}{J.~Phys.~A52,~225401~(2019)}},
\texttt{\arxivref{1812.07287}{arxiv:1812.07287}}.

\bibitem{2004Tolstoy}
V.~N.~Tolstoy,
\textit{``{Chains of extended Jordanian twists for Lie superalgebras}''},
\textsf{ArXiv~Mathematics~e-prints~,~~(2004)},
\texttt{\arxivref{math/0402433}{math/0402433}}.

\bibitem{Patera:1973yn}
J.~Patera, P.~Winternitz and H.~Zassenhaus,
\textit{``{The Maximal Solvable Subgroups Of The $Su(p,q)$ Groups And All
  Subgroups Of $Su(2,1)$}''},
\textsf{\doiref{10.1063/1.1666820}{J.~Math.~Phys.~15,~1378~(1974)}}.

\bibitem{Patera1973b}
J.~Patera, P.~Winternitz and H.~Zassenhaus,
\textit{``The maximal solvable subgroups of SO(p,q) groups''},
\textsf{\doiref{http://dx.doi.org/10.1063/1.1666559}{Journal~of~Mathematical~Physics~15,~1932~(1974)}},
\href{http://scitation.aip.org/content/aip/journal/jmp/15/11/10.1063/1.1666559}{\texttt{http://scitation.aip.org/content/aip/journal/jmp/15/11/10.1063/1.1666559}}.

\bibitem{Hoare:2016wsk}
B.~Hoare and A.~A.~Tseytlin,
\textit{``{Homogeneous Yang-Baxter deformations as non-abelian duals of the
  AdS$_5$ sigma-model}''},
\textsf{\doiref{10.1088/1751-8113/49/49/494001}{J.~Phys.~A49,~494001~(2016)}},
\texttt{\arxivref{1609.02550}{arxiv:1609.02550}}.

\bibitem{Borsato:2016pas}
R.~Borsato and L.~Wulff,
\textit{``{Integrable Deformations of $T$-Dual $\sigma$ Models}''},
\textsf{\doiref{10.1103/PhysRevLett.117.251602}{Phys.~Rev.~Lett.~117,~251602~(2016)}},
\texttt{\arxivref{1609.09834}{arxiv:1609.09834}}.

\bibitem{vanTongeren:2015uha}
S.~J.~van~Tongeren,
\textit{``{Yang–Baxter deformations, AdS/CFT, and twist-noncommutative gauge
  theory}''},
\textsf{\doiref{10.1016/j.nuclphysb.2016.01.012}{Nucl.~Phys.~B904,~148~(2016)}},
\texttt{\arxivref{1506.01023}{arxiv:1506.01023}}.

\bibitem{vanTongeren:2016eeb}
S.~J.~van~Tongeren,
\textit{``{Almost abelian twists and AdS/CFT}''},
\textsf{\doiref{10.1016/j.physletb.2016.12.002}{Phys.~Lett.~B765,~344~(2017)}},
\texttt{\arxivref{1610.05677}{arxiv:1610.05677}}.

\bibitem{Araujo:2017jkb}
T.~Araujo, I.~Bakhmatov, E.~O.~Colg\'{a}in, J.~Sakamoto, M.~M.~Sheikh-Jabbari
  and K.~Yoshida,
\textit{``{Yang-Baxter $\sigma$-models, conformal twists, and noncommutative
  Yang-Mills theory}''},
\textsf{\doiref{10.1103/PhysRevD.95.105006}{Phys.~Rev.~D95,~105006~(2017)}},
\texttt{\arxivref{1702.02861}{arxiv:1702.02861}}.

\bibitem{Araujo:2017jap}
T.~Araujo, I.~Bakhmatov, E.~O.~Colg\'ain, J.-i.~Sakamoto, M.~M.~Sheikh-Jabbari
  and K.~Yoshida,
\textit{``{Conformal twists, Yang\textendash{}Baxter \ensuremath{\sigma}-models
  \& holographic noncommutativity}''},
\textsf{\doiref{10.1088/1751-8121/aac195}{J.~Phys.~A~51,~235401~(2018)}},
\texttt{\arxivref{1705.02063}{arxiv:1705.02063}}.

\bibitem{Beisert:2005tm}
N.~Beisert,
\textit{``{The su$(2|2)$ dynamic S-matrix}''},
\textsf{\doiref{10.4310/ATMP.2008.v12.n5.a1}{Adv.Theor.Math.Phys.~12,~945~(2008)}},
\texttt{\arxivref{hep-th/0511082}{hep-th/0511082}}.

\bibitem{Minahan:2010js}
J.~A.~Minahan,
\textit{``{Review of AdS/CFT Integrability, Chapter I.1: Spin Chains in N=4
  Super Yang-Mills}''},
\textsf{\doiref{10.1007/s11005-011-0522-9}{Lett.~Math.~Phys.~99,~33~(2012)}},
\texttt{\arxivref{1012.3983}{arxiv:1012.3983}}.

\bibitem{Arutyunov:2009ga}
G.~Arutyunov and S.~Frolov,
\textit{``{Foundations of the AdS$_5 \times$S$^5$ Superstring. Part I}''},
\textsf{\doiref{10.1088/1751-8113/42/25/254003}{J.Phys.~A42,~254003~(2009)}},
\texttt{\arxivref{0901.4937}{arxiv:0901.4937}}.

\end{thebibliography}

\end{document}